\documentclass[10pt,journal]{IEEEtran}
\usepackage{cite}
\usepackage{amsmath,amssymb,amsfonts,amsthm}
\usepackage{graphicx}
\usepackage{textcomp}
\usepackage{xcolor}
\usepackage{balance}
\usepackage{geometry}
\usepackage[ruled,linesnumbered,vlined]{algorithm2e}
\usepackage{booktabs}
\usepackage{subcaption}
\usepackage{url}
\usepackage{hyperref}
\usepackage{array}  
\usepackage{booktabs} 
\usepackage{tabularx}
\usepackage{algpseudocode}
\usepackage{algorithm2e}
\usepackage{listings}
\usepackage{fancybox}
\usepackage{caption}

\newtheorem{theorem}{Theorem}
\newtheorem{assumption}{Assumption}
\newtheorem{proposition}{Proposition}

\newtheorem{corollary}{Corollary}
\newtheorem{lemma}{Lemma}

\lstdefinestyle{custompython}{
    language=Python,
    basicstyle=\ttfamily\small, 
    keywordstyle=\color{blue}, 
    commentstyle=\color{green!50!black}, 
    stringstyle=\color{orange}, 
    numbers=left, 
    numberstyle=\scriptsize\ttfamily,
    stepnumber=1, 
    numbersep=8pt, 
    breaklines=true, 
    backgroundcolor=\color{white}, 
    showspaces=false, 
    showstringspaces=false, 
    showtabs=false, 
    tabsize=4, 
    frame=topbottom,
    framerule=0.8pt, 
    framesep=5pt,
    aboveskip=-4mm, 
    framexleftmargin=1.5em, 
    xleftmargin=1.5em 
}

\begin{document}

\title{Learning In Chaos: Efficient Autoscaling and Self-Healing for Multi-Party Distributed Training}

\author{Wenjiao Feng, Rongxing Xiao, Zonghang Li\textsuperscript{\dag}, Hongfang Yu, Gang Sun, Long Luo,\\ Mohsen Guizani,~\IEEEmembership{Fellow,~IEEE}, Qirong Ho, and Steve Liu,~\IEEEmembership{Fellow,~IEEE}

\thanks{This work was supported in part by the National Natural Science Foundation of China (62394324) and Young Elite Scientists Sponsorship Program by CAST (2022QNRC001).}
\thanks{
W. Feng, R. Xiao, H. Yu, G. Sun, and L. Luo are with the School of Information and Communication Engineering, University of Electronic Science and Technology of China (UESTC), Chengdu, 611731, China (email: \{fwj0612, yzzxrx\}@gmail.com, \{yuhf, gangsun, llong\}@uestc.edu.cn).}
\thanks{Z. Li, M. Guizani, Q. Ho, and S. Liu are with the Department of Machine Learning, Mohamed bin Zayed University of Artificial Intelligence (MBZUAI), Building 1B, Masdar City, Abu Dhabi, United Arab Emirates  (email: \{zonghang.li, mohsen.guizani, qirong.ho, steve.liu\}@mbzuai.ac.ae).
}
\thanks{Corresponding authors are Z. Li and H. Yu.}
}

\maketitle

\begin{abstract}
Node and link churn in multi-party, cross-region clusters over wide-area networks (WANs) often disrupts distributed training. However, checkpoint-based recovery and cloud-centric autoscaling react slowly and assume centralized control, which is misaligned with the self-governed setup where institutions can freely join and leave. This paper proposes Chaos, a multi-party distributed training system with self-healing and autoscaling, enabling robust and elastic training under churn. It speeds up autoscaling via multi-neighbor state replication and model sharding. We formalize the sharding and assignment as a MINLP that captures WAN heterogeneity, and reduce it to a tractable MILP by analyzing its monotonicity on a divisibility chain. By establishing an equivalence, we derive a greedy algorithm that follows optimality rules and yields the optimal solution in polynomial time. Chaos uses a cluster monitor to track resource and topology changes, and handles scaling events through peer negotiation protocols, enabling fully self-governed autoscaling among institutions. Experiments show that Chaos has substantially lower scale-out delay than Pollux, Elan, and Autoscaling, and handles scale-in, connect-link, and disconnect-link events within 20ms. It also delivers the lowest idle time, showing superior resource use and scalability as the cluster grows.
\end{abstract}

\section{Introduction}
Vast and critical datasets lie fragmented across individual institutions, creating barriers to innovation, especially in the medical and biotechnology industries. Cross-institution collaboration has emerged as an effective path to overcome these silos. In 2020, NVIDIA launched a federated learning (FL) project across 20 hospitals worldwide and developed an AI system to predict whether a patient with COVID-19 would require supplemental oxygen \cite{dayan2021federated}. ADOPS used FL across breast imaging reports from 7 clinical institutions worldwide and developed a system for breast density classification \cite{roth2020federated}. In 2021, NVIDIA, UCLA, SUNY Upstate, and the NCI collaborated to develop an accurate prostate cancer segmentation model \cite{sarma2021federated}. In 2022, NVIDIA and several U.S. universities used data from 42 hospitals across 5 healthcare systems in the U.S. and Europe to improve chest X-ray–based AI models for COVID-19 diagnosis \cite{peng2022evaluation}. Intel and Penn Medicine coordinated a federation of 71 medical and research institutions worldwide, achieving significant gains in brain tumor recognition \cite{pati2022federated}. In 2023, Owkin, NVIDIA, KCL, and over a dozen pharmaceutical partners used FL to develop an AI model on the world's largest collaborative drug compound dataset \cite{heyndrickx2023melloddy}. Even in the new era of large models, collaborating multi-party data to apply better parameter-efficient fine-tuning on foundation models remains a major focus \cite{zhang2023fedpetuning,bai2024federated,wang2024flora,yan2025federated,singhal2025fedex,sun2024improving}.

\begin{figure*}
    \centering
    \includegraphics[width=\linewidth]{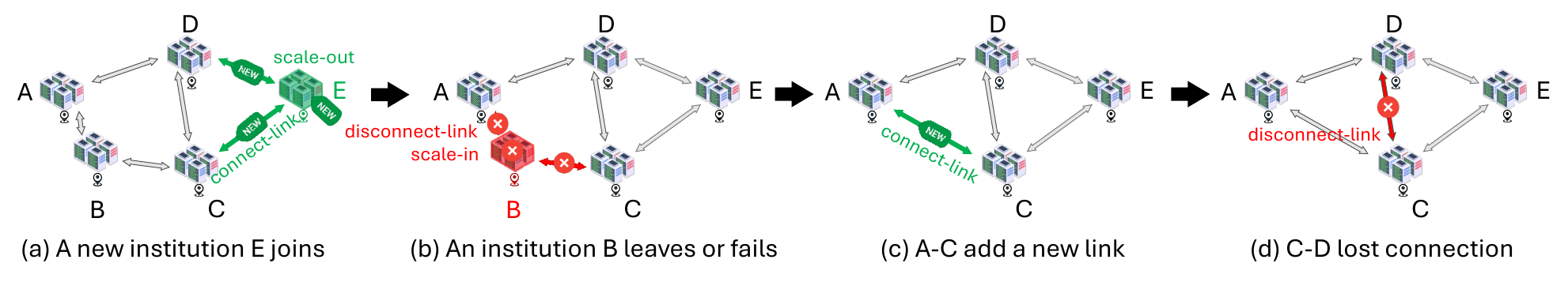}
    \caption{A multi-party cluster under churn: (a) node joining, (b) node leaving, (c) link establishment, and (d) link loss.}
    \label{fig:background}
\end{figure*}

These successful cases \cite{dayan2021federated,roth2020federated,sarma2021federated,peng2022evaluation,pati2022federated,heyndrickx2023melloddy} rely on heavy offline negotiations, where agreements must be signed in person and all institutions must synchronize to start a training task. This rigid process limits generalization: once joined, institutions cannot leave, and outsiders cannot join. To move beyond these constraints, a public and online coordination platform is essential. Such a platform should allow institutions to join or leave freely, as illustrated in Fig. \ref{fig:background}. However, this openness and flexibility also raise \textit{new technical demands: the distributed system must support efficient self-healing and autoscaling to sustain robust training.}
\begin{itemize}
    \item[(a)] \textit{Need for self-healing:} Institutions may go offline due to occasional device, power, network, or software failures. If the system cannot automatically detect the issue, reconfigure, and resume training, the entire system will fail.
    \item[(b)] \textit{Need for autoscaling:} Institutions can participate in training during off-peak business hours and withdraw during peak hours or system maintenance. Without scaling out, new institutions cannot join; without scaling in, existing participants cannot leave.
\end{itemize}

We abstract each institution as a node connected by WAN links, forming a logical overlay. Node or link churn alters this overlay's topology and capacity, and interrupts distributed training. A common fix to handle unexpected node dropouts is checkpoint-restart: save model checkpoints, fix the issue, then roll back and restart training from checkpoints \cite{he2023unicron,maurya2024datastates,mohan2021checkfreq,gupta2024just,eisenman2022check,wang2023gemini}. This stop-resume method also applies to autoscaling: stop training when nodes join or leave, reinit the cluster, and restart training from the last checkpoint \cite{qiao2021pollux,zhang2024rubick,peng2018optimus,chen2023deepboot,lian2025universal}. This process is time- and resource-costly due to heavy disk I/O or network transfer from frequent checkpointing. For example, checkpoint-restart takes $\sim$2min in Pollux \cite{qiao2021pollux} and $\sim$1min in Rubick \cite{zhang2024rubick} and UCP \cite{lian2025universal}, preventing them from frequent scaling in/out. To address this, stop-free autoscaling \cite{ma2019paddlepaddle,zhou2023elasticdl,peng2021dl2,wang2021EPS,wu2021EDL,hwang2021CoDDL,xie2020elan,or2020resource} lets new nodes synchronize training states from running nodes, without restarts or checkpoints. This enables continuous training over cross-region clusters with churn. However, they are still slow in practice. When a new node joins, it must fetch the latest training state from existing nodes, such as model weights. This time cost is referred to as \textit{state replication delay}. In data centers, high-speed LANs keep this delay minimal. Across WANs, however, limited bandwidth can stretch the delay to several minutes, especially when the new node is geographically distant. Such high delays make autoscaling impractical for high-churn clusters. Thus, state replication should be optimized to adapt to node and link changes rapidly.

It should be noted that, unlike GPU scheduling systems in HPC (e.g., Pollux \cite{qiao2021pollux}, Rubick \cite{zhang2024rubick}, and \cite{peng2018optimus,chen2023deepboot,ma2019paddlepaddle,zhou2023elasticdl,peng2021dl2,wang2021EPS,wu2021EDL,hwang2021CoDDL,xie2020elan,or2020resource,lian2025universal}), which have global control over the cluster and can decide to add or remove nodes/GPUs during training for proactive scaling, our setting targets multiple self-governed participants. These participants are not centrally controlled: they join and leave freely, and can also drop out unexpectedly during training. This implies that \textit{autoscaling is triggered by each participating node rather than by a central ``GPU scheduler".} As a result:
\begin{itemize}
    \item[(a)] \textit{Autoscaling occurs more frequently,} since node arrivals and departures are no longer batch operations but individual events. This places stricter requirements on the latency of scale-in and scale-out.
    \item[(b)] \textit{Autoscaling and self-healing can affect model quality.} The addition or removal of nodes corresponds to the injection or withdrawal of training data. Thus, the impact of scale-in and scale-out on model quality should be investigated.
\end{itemize}

This paper proposes Chaos, a multi-party distributed training system with self-healing and autoscaling, enabling robust and elastic training under churn. With this system, institutions can freely join or leave during training and alter their connections. Node or link join, leave, or failure initiates scaling events, then neighbors run peer negotiation protocols to perform \textit{scale-in}, \textit{scale-out}, \textit{connect-link}, and \textit{disconnect-link}. These protocols remove faulty nodes or links, fix communication rules, add or remove participants, and resume training in a fully self-governed manner. To speed up autoscaling, we propose multi-neighbor state replication with model sharding. With this mechanism, a new node can fetch the latest state in parallel from multiple neighbors while balancing their transmission loads to minimize scale-out delay. We set up a cluster with 6–12 parties, connect them with a constrained, heterogeneous WAN, and simulate node joins and exits. Compared to Pollux \cite{qiao2021pollux}, Elan \cite{xie2020elan}, and Autoscaling \cite{or2020resource}, Chaos handles self-healing and autoscaling more efficiently, with scale-out delays between 0.5-5 seconds and others within 20ms. Our main contributions are summarized as follows:

\begin{itemize}
    \item We propose a multi-party distributed training system with self-healing and autoscaling that can more efficiently handle node and link joins, exits, and failures.
    \item We formulate the multi-neighbor state replication problem with model sharding and assignment, derive a greedy solution, and prove its optimality. This yields a network-aware strategy that minimizes replication time.
    \item We propose a cluster monitor that tracks resource and topology changes, along with peer negotiation protocols that enable self-governed autoscaling.
    \item We implement Chaos and demonstrate that it achieves substantially lower scale-out delay than Pollux, Elan, and Autoscaling, with high resource efficiency and scalability.
\end{itemize}
\section{Related Work and Motivations}
In this section, we first review related work on self-healing and autoscaling in distributed machine learning (DML), then explain how our work differs, outline our focus, and present the motivation behind the design of Chaos.

\subsection{Reviews on Self-healing and Autoscaling}
\textit{Self-healing.}
In DML, worker nodes should synchronize model weights frequently, but in chaotic WANs, frequent node and link failures block synchronization and freeze training. Checkpointing helps recover from such failures by periodically saving the training state, so the training can restart from the last checkpoint \cite{he2023unicron}. DataStates-LLM \cite{maurya2024datastates} and CheckFreq \cite{mohan2021checkfreq} reduce training stalls by overlapping computation and checkpoint I/O. JIT Checkpointing \cite{gupta2024just} creates a remote checkpoint only after failures. Check-N-Run \cite{eisenman2022check} combines differential checkpoints with quantization and decoupled background persistence to cut checkpoint size and bandwidth without hurting accuracy. Gemini \cite{wang2023gemini} uses remote CPU memories and high-speed networks to achieve fast checkpointing.

\textit{Autoscaling.}
Checkpointing is also a common approach to autoscaling, known as ``\textbf{stop-resume}" \cite{wu2021EDL}: it saves a checkpoint, pauses training, resizes the cluster, then restarts from the latest checkpoint. To reduce the checkpoint-restart overhead, Optimus \cite{peng2018optimus} reconfigures GPU resources at a fixed ten-minute interval. Pollux \cite{qiao2021pollux} executes reconfiguration and saves checkpoints when cluster size and topology have to change. DeepBoot \cite{chen2023deepboot} uses auto-fast elastic to improve Pollux's cold restarts into fast checkpoint I/O and light communication reconfiguration, thus reducing rescheduling delay. Rubick \cite{zhang2024rubick} performs cost-aware reconfiguration: executes reconfiguration when the restart overhead stays below an empirical threshold. Universal Checkpointing \cite{lian2025universal} reduces reconfiguration time with parallel I/O, high-bandwidth broadcasting, and lazy execution. Nevertheless, they still take more than one minute.

\textbf{Stop-free} autoscaling offers a smoother alternative. When a new node joins, it fetches the training state from existing nodes' memory, so there is no need to restart or load/write checkpoints from/to a slow disk. In parameter server setups, PaddleEDL \cite{ma2019paddlepaddle} and ElasticDL \cite{zhou2023elasticdl} use asynchronous training to realize stop-free scaling. As workers run independently, their join and exit don't disrupt others, and new workers fetch the latest model weights from the parameter server. However, they fall back to stop-resume in synchronous mode. DL$^2$ \cite{peng2021dl2} and EPS \cite{wang2021EPS} enable stop-free scaling in synchronous mode, and also require new nodes to fetch the training state from parameter servers. EDL \cite{wu2021EDL} and CoDDL \cite{hwang2021CoDDL}, on the other hand, fetch from a single worker node. Autoscaling \cite{or2020resource} goes further by slicing the training state across multiple nodes and enabling parallel fetching for faster replication. Elan \cite{xie2020elan}, in contrast, explores a different path by having the new node fetch from the neighbor with the highest available bandwidth.

\subsection{Differences and Focus of This Paper}
The prior works focus on optimizing reconfiguration policies (e.g., adjusting execution plans and GPU allocations) via a centralized GPU scheduler, but deploy new configurations via simple checkpoint-restart or naive stop-free approaches, leading to minute-scale delays during scale-in and scale-out. Instead, \textit{Chaos targets minimizing these scale-in and scale-out delays, and in our scenario:} 
\begin{itemize}
    \item \textit{A centralized node/GPU scheduler doesn't exist} as scaling events are triggered independently by each self-governed institution, which can join or leave freely. As a result, \textit{nodes join and leave one at a time, rather than in batches.}
    \item \textit{During scaling, training data joins or leaves with the institution node,} unlike in HPC, where data can be migrated to other nodes. This may affect model quality.
\end{itemize}
Therefore, cluster scaling will occur more often, demanding lower scaling latency. These differences distinguish Chaos from prior work.

Gossip FL (peer-to-peer, decentralized) also supports self-healing and autoscaling. However, recent studies have shown that gossip protocols often fail to converge under high churn \cite{vikstrom2021comparing}, where nodes frequently join and leave, while centralized FL preserves both convergence and accuracy \cite{chang2024federated}. Therefore, we focus on the centralized setup. Please note that while Chaos uses a centralized architecture for the data plane (e.g., the transmission of model updates), our scaling-delay optimization in the control plane follows a decentralized design, where neighbor institutions can exchange control messages directly. This decoupling of data and control planes makes this work complementary to existing optimizations. For example, we can correct model updates on the server side to restore training stability under high churn \cite{sun2023mimic}; adjust worker workload to mitigate stragglers \cite{li2020esync}; schedule data-plane traffic to accelerate model synchronization \cite{li2024accelerating}; and select training data or nodes to reduce non-IID impact \cite{li2022data}. These techniques can be directly applied, but are not the focus of this paper.

% \textcolor{red}{In existing gossip-based approaches, Moshpit SGD\cite{ryabinin2021moshpit} downloads parameters from a single neighbor, while FedLay \cite{hua2024towards} downloads full parameters from multiple neighbors, resulting in traffic loads several times higher than centralized PS architectures and exponentially increased state synchronization costs, further hindering cluster autoscaling and self-healing. Therefore, achieving fast scaling remains a core challenge for edge training.}

\begin{figure}[t]
\centerline{\includegraphics[width=\linewidth]{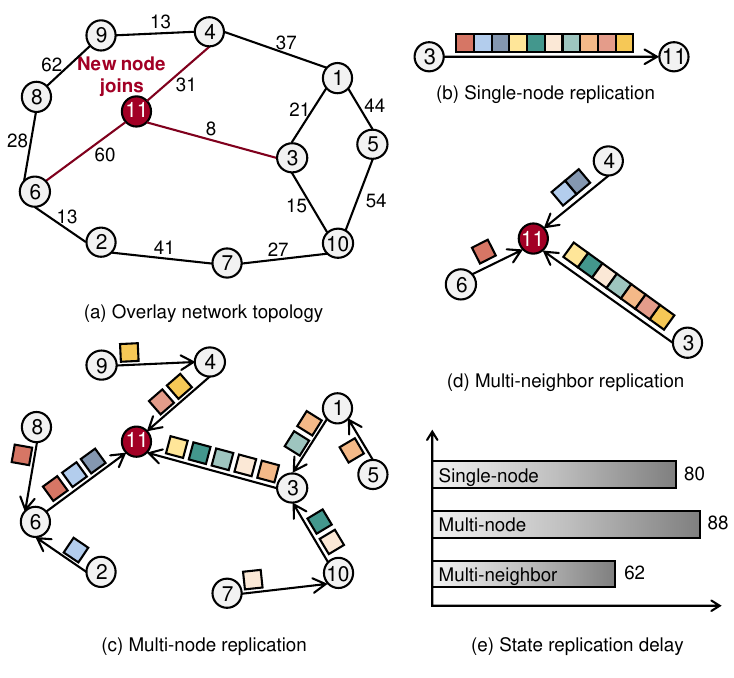}}
\caption{An example of (a) the overlay network topology, (b) single-node replication, (c) multi-node replication, and (d) multi-neighbor replication. Figure (e) compares their state replication delays. The number on each edge indicates the delay incurred to transfer one unit of data through that link. Suppose there are 10 data units to transmit.}
\label{fig:Motivation1}
\end{figure}

\subsection{Motivations}\label{sec:motivation}
\textit{Question A (QA): From which node(s) should a new node fetch to speed up state replication?}

Scale-out delay is the most significant among scaling actions, and in the worst case, it can reach tens of seconds. This delay depends on which node(s) the new node fetches the latest training state from.

In the overlay network topology shown in Fig. \ref{fig:Motivation1}a, each node represents an institution. Prior work \cite{wu2021EDL,hwang2021CoDDL,xie2020elan} fetches the training state from a single worker node (see Fig. \ref{fig:Motivation1}b), which may or may not be a neighbor. Following the idea in \cite{xie2020elan}, when a new node 11 joins, it fetches the training state from its highest-bandwidth neighbor 3, but sending the entire state over a single link creates a bottleneck: state replication still takes 80 time units. 

Multi-source solutions \cite{ma2019paddlepaddle,zhou2023elasticdl,peng2021dl2,wang2021EPS,or2020resource} fetch different shards of the training state in parallel from multiple nodes (Fig. \ref{fig:Motivation1}c), which can be parameter servers or worker nodes. Even with multi-path transfers, the new node may be far from the sources, requiring multiple hops and sometimes crossing bottleneck links. In Fig. \ref{fig:Motivation1}c, this can cause delays by up to 88 time units. This approach is used by default in parameter server systems but often suffers from unpredictable efficiency.

This leads to an insight: \textit{can a new node fetch different shards of the training state from multiple neighbors? Yes!} In synchronous training, parameter servers and workers maintain identical model weights, so a new node can obtain the latest model by fetching at the right time. To handle heterogeneous WAN bandwidth, the model should be appropriately sharded to balance the load across different transfer paths. Fig. \ref{fig:Motivation1}d shows an example: the new node 11 fetches most model shards from high-bandwidth neighbor 3, while also fetching fewer shards in parallel from lower-bandwidth nodes 4 and 6. In this way, the state replication delay is reduced to 62 time units. This calls for \textit{a smart shard scheduler that minimizes the completion time to fetch from neighbors.}

\textit{Question B (QB): How do nodes join or leave a self-governed cluster?}

\begin{figure}[t]
\centerline{\includegraphics[width=\linewidth]{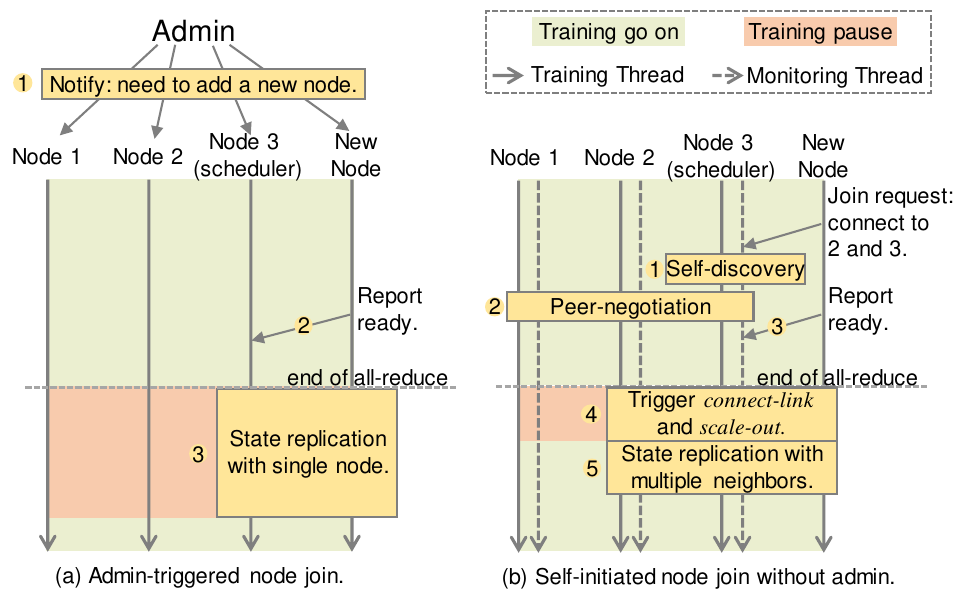}}
\caption{Example of node joins with and w/o a central admin.}
\label{fig:Motivation2}
\end{figure}

Prior work \cite{peng2018optimus,qiao2021pollux,chen2023deepboot,zhang2024rubick,ma2019paddlepaddle,zhou2023elasticdl,peng2021dl2,wang2021EPS,wu2021EDL,hwang2021CoDDL,xie2020elan,or2020resource,lian2025universal} has generally relied on a GPU scheduler to make scaling decisions. This central scheduler holds full control over all nodes and GPUs, so it can add or remove them in batches at set times. In Fig. \ref{fig:Motivation2}a, the admin acts as this scaling scheduler: it sends join commands and authorizes new nodes to participate in the next training round. This paradigm, however, does not apply to self-governed clusters.

In a self-governed cluster, institutions join and leave on their own schedule, and no single party controls the others. Thus, a central ``scaling scheduler" would lack the authority to command or batch participating parties, and joins and leaves must be handled locally, one at a time, through peer negotiation. As shown in Fig. \ref{fig:Motivation2}b, each node runs a lightweight agent to discover neighbor and link changes, negotiate admission, trigger scaling actions (e.g., connect-link and scale-out), and synchronize training state to the new node. Similarly, departures or failures are discovered by neighbors, which then negotiate and execute actions (e.g., disconnect-link, scale-in) to restore training. This requires a \textit{cluster monitor} to detect node and link changes, and \textit{peer negotiation protocols} to coordinate scale-in, scale-out, connect-link, and disconnect-link actions.

To address QA, Section \ref{sec:multi-neighbor-state-replication} formulates the model sharding and assignment problem, introduces a polynomial-time algorithm, and analyzes its optimality and complexity. For QB, Section \ref{sec:cluster-monitor-and-negotiation} presents a cluster monitor and peer negotiation protocols for self-governed scaling. Section \ref{system_design} outlines the system design, and Section \ref{sec_implementation} provides experimental results.
\section{Multi-neighbor State Replication}\label{sec:multi-neighbor-state-replication}
To minimize the state replication delay when a new node joins, this section formulates a model sharding and assignment problem. The original mixed-integer nonlinear program (MINLP) P2 is intractable, so we analyze the monotonicity on a divisibility chain and use its non-decreasing nature to reduce P2 to a 0-1 mixed-integer linear program (0-1 MILP) P3. However, solving P3 with modern MILP solvers remains slow and prone to OOM. We therefore prove that P3 is a special case of MILP, equivalent to a multi-way merge problem, where a simple greedy rule yields the optimal solution. Based on this insight, we design a polynomial-time greedy algorithm and confirm its optimality through experiments.

% \begin{figure}[t]
% \centerline{\includegraphics[width=.85\linewidth]{figures/Characteristics-of-training-states.pdf}}
% \caption{State replication delays for model weights, optimizer states, and other runtime information. The link bandwidth is set to 200 Mbps.}
% \label{fig:Training States}
% \end{figure}

\subsection{Model Sharding and Assignment Problem}
\label{sec_problem_model}
The training state contains tensors of varying sizes, from small ones like convolution kernels and biases to large fully connected weights. Assigning these raw tensors directly to neighbors would cause uneven traffic. To balance the load, we split tensors into equal-sized shards and assign them across neighbors. Here, ``assignment" indicates that each shard is sent by a different neighbor to the new node. If shards are too large, there would be fewer shards and traffic remains uneven; if too small, shard management becomes expensive.

Let $G = (V, E)$ be a weighted graph, with each edge having a weight representing the delay of transmitting one unit of data over that link. Given a DML task with model weights $w$, to achieve a fine-grained control over traffic load, we split $w$ into $K$ shards, each with data size $s$. Assume $s$ can divide the total data size $|w|$, that is, $|w|=Ks$. Let $\mathcal{K}=\{w_1,w_2,\cdots,w_K\}$ be the set of model shards. Now, a new node $v_{\text{new}}$ joins and needs to fetch shards $\mathcal{K}=\{\mathcal{K}_1,\cdots,\mathcal{K}_{|\mathcal{U}|}\}$ from its neighbors $\mathcal{U}$, where each neighbor $u\in\mathcal{U}$ is assigned an non-empty shard subset $\mathcal{K}_u\ne\emptyset,\mathcal{K}_u\subset\mathcal{K}$. These shard subsets are disjoint and their collection covers all shards, i.e., $\bigcup_{u \in \mathcal{U}} \mathcal{K}_u = \mathcal{K}$ and $\mathcal{K}_u \cap \mathcal{K}_v = \emptyset$ for any $u \ne v$.

Let $t^{\text{prop}}_{u\rightarrow v_{\text{new}}}$ be the propagation delay and $t^{\text{trans}}_{u\rightarrow v_{\text{new}}}$ the per-parameter transmission delay from neighbor $u$ to $v_{\text{new}}$. Then, the delay to fetch all shards in the set $\mathcal{K}_u$ is:
\begin{equation}
t_u(s,\mathcal{K}_u) = t^{\text{prop}}_{u \rightarrow v_{\text{new}}} + s \cdot t^{\text{trans}}_{u \rightarrow v_{\text{new}}}\cdot|\mathcal{K}_u|.
\end{equation}

As shown in Fig. \ref{fig:Motivation2}b, state replication starts after the all-reduce, but neighbors finish all-reduce at different times. Let $\tau_u^{\text{sync}}$ be the time when neighbor $u$ finishes all-reduce. Our goal is to minimize the latest completion time among all neighbors finishing their state replication to the new node:
\begin{align}
\min_{s,K,\mathcal{K}_u} \max_{u \in \mathcal{U}} \quad & t_u(s,\mathcal{K}_u) + \tau_u^{\text{sync}}, \\
\text{s.t.} \quad & s \in \mathbb{Z}^+, |\mathcal{K}_u|>0, \forall u\in\mathcal{U}, \\
& K \ge |\mathcal{U}|,~|w|=Ks.
\end{align}

To formally represent $\mathcal{K}_u$, we introduce a binary selection matrix $x_{uj}\in\{0,1\}$, where $u\in\mathcal{U}$ is a neighbor of the new node and $j\in\{1,2,\cdots,K\}$ indexes the model shards. If $x_{uj}=1$, neighbor $u$ sends shard $w_j$ to the new node. Thus, the problem can be reformulated as \textbf{P1}:
\begin{align}
\min_{s,K,x_{uj}} \max_{u \in \mathcal{U}} \quad & t_u(s,x_{uj}) + \tau_u^{\text{sync}}, \\
\text{s.t.} \quad & t_u(s,x_{uj})=t^{\text{prop}}_{u \rightarrow v_{\text{new}}} + \sum_{j=1}^K s t^{\text{trans}}_{u \rightarrow v_{\text{new}}} x_{uj}, \\
& \sum_{u\in\mathcal{U}}x_{uj}=1,~\forall j\in\{1,2,\cdots,K\}, \\
& \sum_{j=1}^{K}x_{uj}\ge 1,~\forall u\in\mathcal{U}, \\
& x_{uj}\in\{0,1\},~\forall u\in\mathcal{U},j\in\{1,2,\cdots,K\}, \\
& s \in \mathbb{Z}^+, K \ge |\mathcal{U}|,~|w|=Ks.
\end{align}

This objective involves a max operation, so we introduce an auxiliary variable $\theta$ to bound the completion time across all neighbors. Then the problem is rewritten to \textbf{P2}:
\begin{align}
\min_{s, K, x_{uj}, \theta} \quad & \theta \\
\text{s.t.} \quad 
& t_u(s,x_{uj}) + \tau_u^{\text{sync}} \le \theta,~\forall u\in\mathcal{U}, \\
& t_u(s,x_{uj})=t^{\text{prop}}_{u \rightarrow v_{\text{new}}} + \sum_{j=1}^K s \cdot t^{\text{trans}}_{u \rightarrow v_{\text{new}}}\cdot x_{uj}, \\
& \sum_{u \in \mathcal{U}} x_{uj} = 1,~\forall j \in \{1, 2, \cdots, K\}, \\
& \sum_{j=1}^{K}x_{uj}\ge 1,~\forall u\in\mathcal{U}, \\
& x_{uj} \in \{0, 1\},~\forall u\in\mathcal{U},j\in\{1,2,\cdots,K\}, \\
& s \in \mathbb{Z}^+, K \ge |\mathcal{U}|,~|w|=Ks.
\end{align}

If both $s,K$ are treated as decision variables, the constraints contain nonlinear terms such as $sx_{uj}$ and $Ks$, with $s,K$ being positive integers and $x_{uj}$ binary. This makes P2 a mixed-integer nonlinear program (MINLP). In addition, because the summation indices and constraint sets depend on $K$, the model is not of fixed dimension, making it intractable.

To make the problem solvable, we apply an outer enumeration over $d$ to fix $s=2^d$ in P2, so that each subproblem reduces to a linear 0-1 program with fixed dimension.

\begin{assumption}\label{assumption1}
For any neighbor $u$, its traffic completion time depends only on the total size of its assigned shards, not on $s$ or $K$ separately. In other words, we ignore per-shard extra cost (e.g., packetization, handshake), since small shards can be batched into a single data stream for transmission.
\end{assumption}

\begin{theorem}[Monotonicity on a divisibility chain]\label{theorem:monotonicity}
Let Assumption \ref{assumption1} hold, if $s_1 \mid s_2$ with $s_1 < s_2$, then $\theta^{(s_1)}\ \le\ \theta^{(s_2)}$.
\end{theorem}
\begin{proof}
Let $q:= s_2/s_1 \in \mathbb Z_{>0}$. Take any feasible assignment $x^{(2)}=\{x_{uj}^{(2)}\}$ for $P2(s_2)$ and let $n_u^{(2)} := \sum_{j=1}^{K(s_2)} x_{uj}^{(2)}$ be the number of shards assigned to neighbor $u$ under $x^{(2)}$. Then, construct $x^{(1)}=\{x_{u\ell}^{(1)}\}$ for $P2(s_1)$ by splitting each $s_2$-shard into $q$ shards of size $s_1$, and
assigning all of them to the same neighbor as before. Then for every neighbor $u$: 
$$n_u^{(1)} = q\, n_u^{(2)},~s_1\, n_u^{(1)} = s_1 q\, n_u^{(2)} = s_2\, n_u^{(2)}.$$
Hence, each neighbor's completion time is preserved:
$$t_u^{\text{prop}}+\tau_u^{\text{sync}}
+ s_1 t_u^{\text{trans}} n_u^{(1)}
=
t_u^{\text{prop}}+\tau_u^{\text{sync}}
+ s_2 t_u^{\text{trans}} n_u^{(2)}.$$

Coverage is maintained since $K(s_1)=qK(s_2)$ and every $s_2$-shard is replaced by $q$ shards of size $s_1$. Since each neighbor holds at least one shard, $n_u^{(1)}=q n_u^{(2)} \ge 1$ whenever $n_u^{(2)}\ge 1$.
Therefore, $x^{(1)}$ is feasible for $P2(s_1)$ and attains the same
objective value as $x^{(2)}$, which implies $\theta^{(s_1)}\ \le\ \theta^{(s_2)}$.
\end{proof}

\begin{corollary}\label{corollary1}
If the candidate set $\mathcal{S}=\{s_1 < s_2 < \cdots < s_m\}$ satisfies
$s_i \mid s_{i+1}$ for all $i$, then
\[
\theta^{(s_1)} \ \le\ \theta^{(s_2)} \ \le\ \cdots \ \le\ \theta^{(s_m)},
\]
i.e., $\theta^{(s)}$ is non-decreasing along the divisibility chain.
\end{corollary}

According to Corollary \ref{corollary1}, the optimal setup is $s_1$, which could be as small as one element or an atomic size. In practice, over-sharding (e.g., shards of size one) should be avoided, so an appropriate $s_{\min}$ should be set as the atomic size. A practical choice is the size of the smallest vector, such as biases or LayerNorm parameters. Take the Transformer model as an example: these vectors have length $H$, while most weight matrices are shaped $H\times mH$, where $m$ is an integer. Setting $s_{\min} = H$ ensures most tensors are divisible, prevents uneven shards, simplifies shard management, and still provides fine-grained load balancing across neighbors.

Thus, with $s$ and $K$ fixed, problem P2 reduces to P3:
\begin{align}
\min_{x_{uj}, \theta} \quad & \theta \\
\text{s.t.} \quad 
& t_u(x_{uj}) + \tau_u^{\text{sync}} \le \theta,~\forall u\in\mathcal{U}, \label{cons:theta-bound} \\
& t_u(x_{uj})=t^{\text{prop}}_{u \rightarrow v_{\text{new}}} + s \cdot t^{\text{trans}}_{u \rightarrow v_{\text{new}}} \cdot \sum_{j=1}^K x_{uj}, \label{cons:t_u} \\
& \sum_{u \in \mathcal{U}} x_{uj} = 1,~\forall j \in \{1, 2, \cdots, K\}, \\
& \sum_{j=1}^{K}x_{uj}\ge 1,~\forall u\in\mathcal{U}, \\
& x_{uj} \in \{0, 1\},~\forall u\in\mathcal{U},j\in\{1,2,\cdots,K\}.
\end{align}

% Before that, let's fix $x_{uj}$ and optimize only $s$, the problem P2 becomes:
% \begin{equation}
% \min_s \max_{u\in\mathcal{U}}~\alpha_u \cdot s + \beta_u,~\text{s.t.}~s \in \mathbb{Z}^+,
% \end{equation}
% where $\alpha_u>0, \beta_u>0$ are constants. Since the objective is a max over linear functions, the overall objective is strictly increasing with respect to $s$. Although $\theta$ in P2 is not strictly monotonic, it often exhibits a quasi-monotonic trend in practice \cite{boyd2004convex}, making binary search an effective way to quickly find a near-optimal $s$, as described in Algorithm \ref{alg:OptimalPartition}.

% Then, we can apply external binary search to fix $s$ and focus on optimizing $x_{uj}$. The problem P2 is then rewritten to \textbf{P3}:

P3 is a standard 0-1 MILP solvable by modern solvers, but for large models with $K\in(10^6,10^8)$ shards and complex networks (many neighbors $|\mathcal{U}|$), solving it can still take a long time or even run out of memory, because it has $K|\mathcal{U}|+1$ decision variables. Each time a new node joins, the solver must be rerun, so millisecond-level latency is required each run. Most MILP solvers cannot meet such strict time and memory demands. Thus, a fast solver is needed.

% \begin{algorithm}[t]
% \small
% \caption{$\mathsf{Shard\ Assignment\ with\ Binary~Search~(P2)}$}
% \label{alg:OptimalPartition}
% \KwIn{
%     Training states $w$, neighbor set $\mathcal{U}$, new node $v_{\text{new}}$.
% }
% \KwOut{
%     Shard assignment $\{\mathcal{K}_u^*,~\forall u\in\mathcal{U}\}$, Shard size $s^*$.
% }

% Initialize $s_{l} \gets \text{min\_layer\_size}(w)$, $s_{h} \gets \text{max\_layer\_size}(w)$; 

% Initialize the objective value $\theta^* \gets +\infty$;

% \While{$s_{l} \leq s_{h}$}{
%     Set $s \gets \lfloor \frac{s_{l} + s_{h}}{2} \rfloor$;

%     Split tensors in $w$ into a set of shards $\mathcal{K}$, each of size $s$;
    
%     Solve \textbf{P3} with given $\mathcal{K}$ and $s$: \\
%     $\{\mathcal{K}_u,\forall u\} \gets \mathsf{Greedy\ Shard\ Assignment}(\mathcal{K},\mathcal{U},s, v_{\text{new}})$; 
    
%     % Compute delays based on allocation $\mathcal{A}$ and network parameters
%     Calculate objective value $\theta$ in \textbf{P2} with given $\mathcal{K}_u$ and $s$;
    
%     \If{$\theta < \theta^*$}{
%         $\theta^* \leftarrow \theta$, $\mathcal{K}_u^* \gets \mathcal{K}_u$, $s^* \gets s$, \\
%         $s_{h} \gets s - 1$\;
%     }
%     \Else{
%         $s_{l} \gets s + 1$\;
%     }
% }
% \Return{$\{\mathcal{K}_u^*,~\forall u\in\mathcal{U}\},s^*$;}
% \end{algorithm}
\begin{algorithm}[t]
\small
\caption{$\mathsf{Greedy\ Sharding\ and\  Assignment}$}
\label{alg:FeasibleAllocation}
\KwIn{
Model shard set $\mathcal{K}$, neighbor set $\mathcal{U}$, shard size $s$, new node $v_{\text{new}}$, propagation delay $t_{u\rightarrow v_{\text{new}}}^{\text{prop}}$, transmission delay $t_{u\rightarrow v_{\text{new}}}^{\text{trans}}$, 
synchronization delay $\tau_u^{\text{sync}}$ for each $u\in\mathcal{U}$. 
}
\KwOut{
Shard assignment $\{\mathcal{K}_u\}$ for each $u\in\mathcal{U}$.
}

Initialize $\mathcal K_u \gets \emptyset$, $r_u \gets t_{u\rightarrow v_{\text{new}}}^{\text{prop}} + \tau_u^{\text{sync}}$, $p_u \gets s \cdot t_{u\rightarrow v_{\text{new}}}^{\text{trans}}$, and $l_u \gets r_u$;

\ForEach{neighbor $u\in\mathcal U$}{
Pop a shard $w_k$ from $\mathcal{K}$ and append it to $\mathcal{K}_u$;

Update makespan $l_u \gets l_u + p_u$;
}

\While{$\mathcal{K}$ not empty}{
Select the neighbor $u$ with least next-finish time: $u \gets \arg\min_{u\in\mathcal{U}} (l_u + p_u)$\;

Pop a shard $w_k$ from $\mathcal{K}$ and append it to $\mathcal{K}_u$;

Update makespan $l_u \gets l_u + p_u$;
}

\Return{$\{\mathcal{K}_u,~\forall u\in\mathcal{U}\}$};
\end{algorithm}

\subsection{Algorithm Design}\label{sec_state_replication_mechanism}
Through an equivalent transformation, we find that P3 is equivalent to a multi-way merge problem with sorted sequences, where a greedy solution is optimal. Based on the mathematical structure of the problem, we design a greedy sharding and assignment algorithm. Both theory and experiments confirm that it yields optimal solutions and can be solved in polynomial time.

\begin{proposition}[Problem Equivalence]\label{prop:equivalence}
Let $s, K$ be fixed and constants $r_u = t^{\text{prop}}_{u} + \tau_u^{\text{sync}}, p_u=st^{\text{trans}}_{u}$, and define aggregated counts 
\begin{equation}
n_u = \sum_{j=1}^K x_{uj},~\sum_{u\in\mathcal{U}} n_u = K,
\end{equation}
Then P3 is equivalent to P4:
\begin{equation}
\min_{\{n_u|n_u \ge 1\}} \ \max_{u\in\mathcal{U}} \ \bigl(r_u + p_u n_u \bigr)
\quad \text{s.t.} \quad \sum_{u\in\mathcal{U}} n_u = K,
\end{equation}
which is a multi-way merge problem with sorted sequences.
\end{proposition}

\begin{proof}
Assigning one more shard to neighbor $u$ raises its completion time to the next value in the arithmetic progression $$\mathcal T_u=\{r_u+p_u,\ r_u+2p_u,\ r_u+3p_u,\ldots\}.$$ To realize the $j$-th completion at time $r_u+jp_u$, the 1st through $(j-1)$-th completions must have occurred earlier at times $r_u+p_u,\ldots,r_u+(j-1)p_u$. If neighbor $u$ receives $n_u$ shards, the set of its completion times must be the prefix of length $n_u$ of $\mathcal T_u$. Hence, any feasible allocation is the same as selecting exactly $K$ elements from the union $\bigcup_{u\in\mathcal{U}} \mathcal T_u$, with at least one element from each sequence. In this way, the overall makespan equals our goal:
$$\max\{\text{the chosen }K\text{ elements}\}=\max_{u\in\mathcal U}(r_u+p_u n_u).$$
Then the problem becomes: choose $K$ elements from the ordered sequences $\{\mathcal T_u,~\forall u\in\mathcal{U}\}$ (respecting prefix and $n_u\!\ge\!1$) to minimize the maximum selected value. This is exactly a multi-way merge problem with sorted sequences. 
\end{proof}

\begin{lemma}\label{lemma1}
The greedy method provides the optimal solution to the multi-way merge problem with sorted sequences.
\end{lemma}

\begin{proof}
At any step, the best choice is exactly the current head of the sequences (its prefixes already taken). The greedy rule always picks the smallest head. Hence, the first $K$ outputs are the $K$ smallest feasible elements and their maximum is minimal. For example, the min-heap (a heap-based greedy) is the optimal solution to such problems \cite{donald1999art}.
\end{proof}

\begin{figure}[t]
\centering
\includegraphics[width=.8\linewidth]{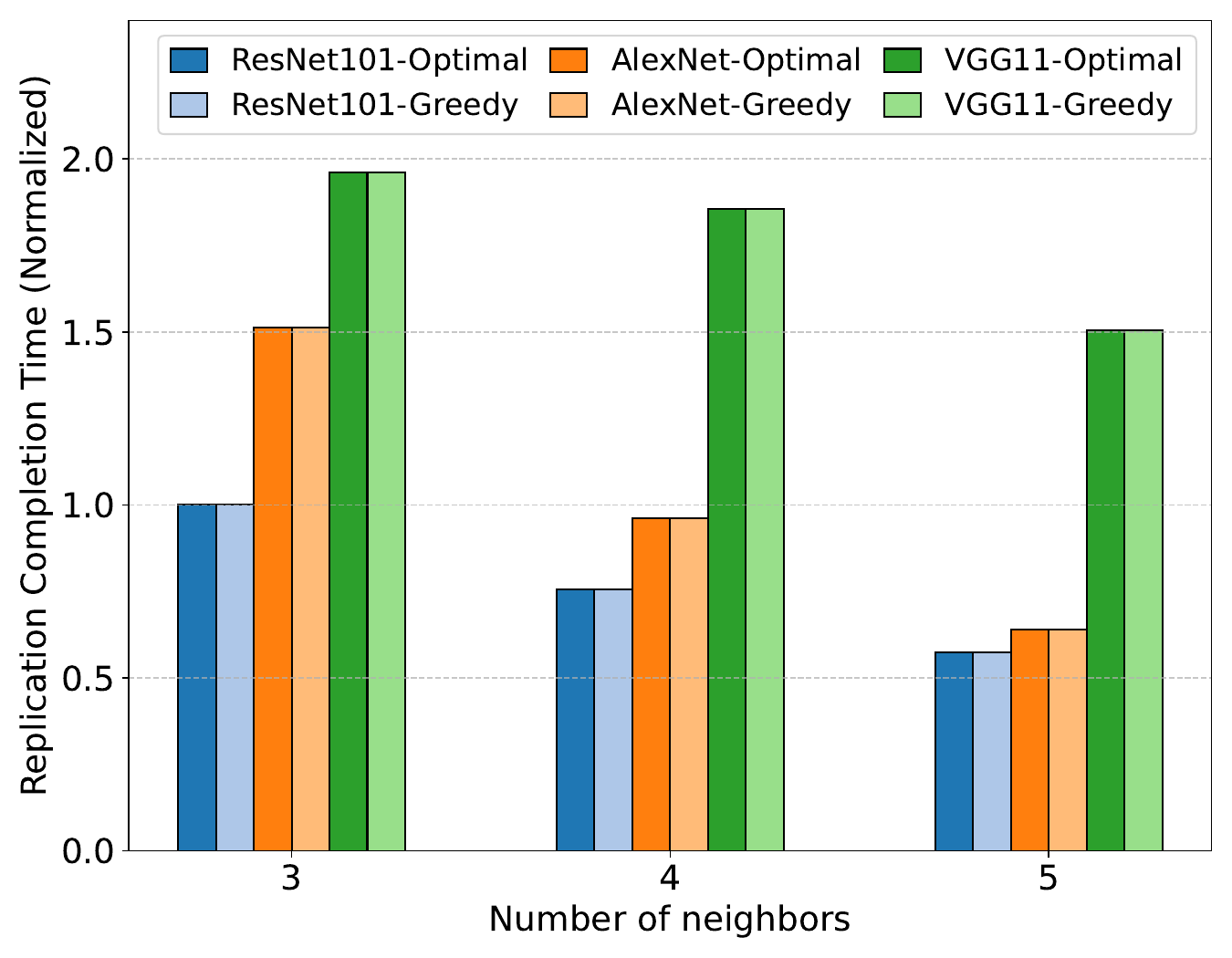}
\caption{Comparison of our greedy solution with the optimal solution on 3 models under 3–5 neighbor nodes.}
\label{fig:greedy-opt}
\end{figure}

\begin{figure*}[t]
\centerline{\includegraphics[width=\linewidth]{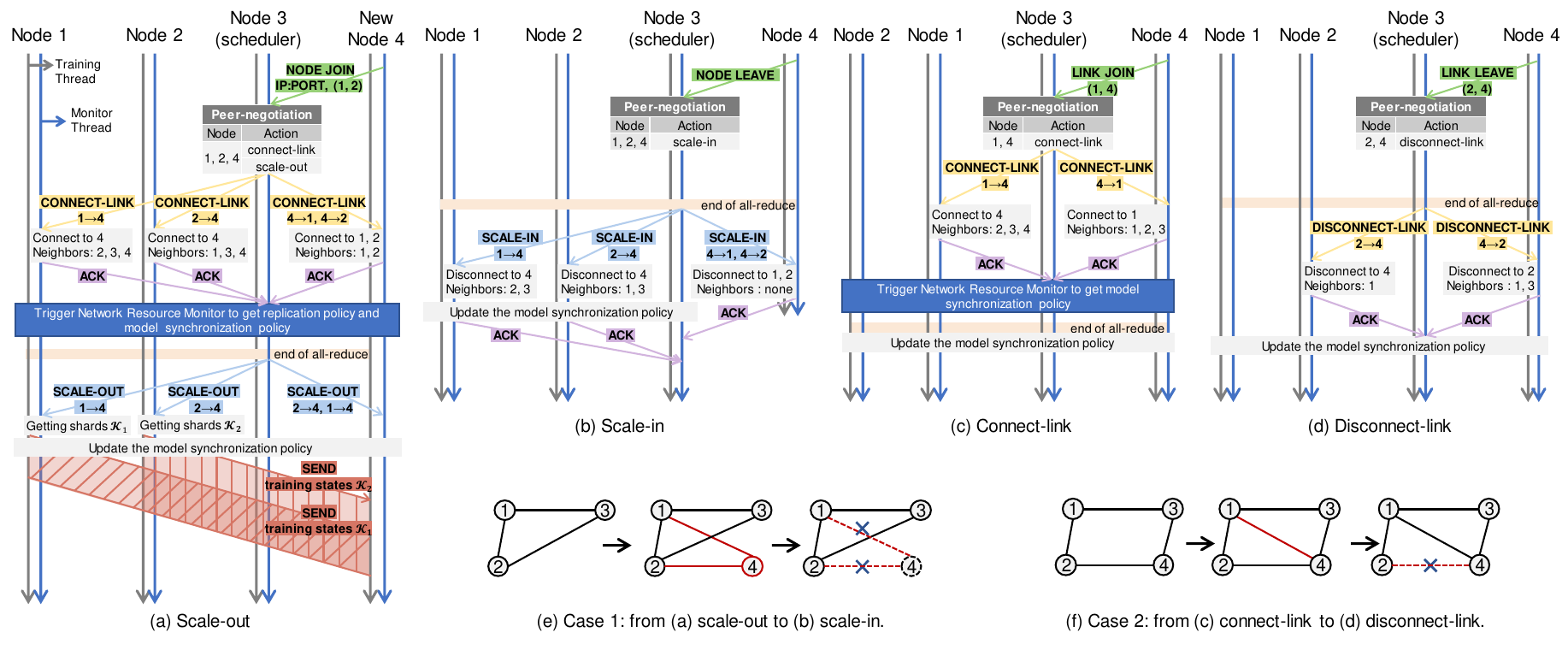}}
\caption{Peer negotiation protocols for scale-out, scale-in, connect-link, and disconnect-link.}
\label{fig:peer-negotiation-protocols}
\end{figure*}

\begin{table*}[t]
\caption{Summary of scaling primitives in Chaos}
\label{tab:commands}
\centering
\begin{tabular}{p{0.1\textwidth} p{0.45\textwidth} p{0.15\textwidth} p{0.10\textwidth}}
\toprule
\textbf{Primitives} & \textbf{Functionality} & \textbf{Event} & \textbf{Delay} \\
\midrule
\textit{Scale-out}        & Add a node, connect it to neighbors, and synchronize training state.                        & Node joins           
            & 500ms-5s \\
\textit{Scale-in}         & Remove a node and terminate its sockets to neighbors. & Node exits or failure     & 20 ms \\
\textit{Connect-link}     & Establish a new socket between two nodes.     & Link joins      
            & 20 ms \\
\textit{Disconnect-link}  & Terminate the socket between two nodes.            & Link exits or failure     & 20 ms \\
\midrule
\end{tabular}
\vspace*{-3mm}  
\end{table*}

Proposition \ref{prop:equivalence} and Lemma \ref{lemma1} motivate us to design a simple but optimal greedy algorithm to solve P3: 
\begin{enumerate}
\item Take the first element from each $\mathcal{T}_u$ to ensure $n_u \ge 1$; 
\item Let $\mathcal H$ be the set of currently available heads;
\item Pick $u=\arg\min_{u\in{\mathcal{U}}} \mathcal H$ and update $\mathcal H$, repeat until $K$ elements are chosen. 
\end{enumerate}

In Algorithm \ref{alg:FeasibleAllocation}, $r_u$ is the one-time propagation and synchronization delay from neighbor $u$, and $p_u$ is the per-shard transmission cost. The variable $l_u$ tracks the current finish time of $u$. First, each neighbor receives one shard to enforce $n_u \ge 1$. Then, for each remaining shard, we select the neighbor with the smallest next-finish time and assign it a shard. This greedy rule is equivalent to choosing the smallest available head in the multi-way merge view. We repeat this greedy rule until all $K$ shards are assigned.

\textbf{Greedy Optimality.} According to Proposition \ref{prop:equivalence} and Lemma \ref{lemma1}, the greedy rule in Algorithm \ref{alg:FeasibleAllocation} should produce the optimal solution. Our preliminary experiment confirms this. As shown in Fig. \ref{fig:greedy-opt}, brute-force search for P4 finds the same minimum.

\textbf{Complexity Analysis.} Algorithm \ref{alg:FeasibleAllocation} scans all $K$ shards and assigns each shard to the least-loaded neighbor among the $m=|\mathcal U|$ candidates, giving complexity $O(mK)$. Given a common setting with $s=2048$, $m=4$, and AlexNet, the solving time is 160ms. However, as noted in Section \ref{sec:overlap}, this latency can be fully overlapped by all-reduce. With a min-heap, the complexity can be reduced to $O((K-m)\log m)$.

Although no new algorithm is proposed, this section shows how an intractable MINLP can be elegantly reduced to a special MILP solvable by a polynomial-time greedy algorithm that yields exact optimal solutions. This has been a significant and practical contribution.
\section{Cluster Monitor and Peer Negotiation}
\label{sec:cluster-monitor-and-negotiation}
In the previous section, we showed how model sharding and assignment can speed up state replication when new nodes join, but two issues remain:
\begin{enumerate}
    \item The algorithm relies on prior knowledge of the cluster topology and resources, like which nodes are neighbors, and the link latency, bandwidth between them.
    \item The algorithm is triggered by cluster changes, such as node joins, exits, and failures, so it relies on timely detection of these events to respond to changes.
\end{enumerate}

In this section, we first introduce a cluster monitor to track resources for scheduling decisions and to detect node and link changes, followed by peer-negotiation protocols to handle self-governed scale-in/out and connect/disconnect-link events.

\subsection{Cluster Monitor}
As stated above, the cluster monitor plays three key roles: tracking the cluster topology, detecting node and link changes, and monitoring network resources.

\textbf{Cluster Topology Monitoring.} 
The primary task of the cluster monitor is to detect the overlay topology (an example is shown in Fig. \ref{fig:Motivation1}a). In Chaos, this is handled by a scheduler, which tracks each node's ID, state (e.g., active, failed, or standby), neighbors, and join time. This helps Algorithm \ref{alg:FeasibleAllocation} identify neighbors $\mathcal{U}$ and spot changes to trigger scaling events.

\textbf{Cluster Event Monitoring.}
In Chaos, scaling events are triggered by node and link changes. The cluster monitor listens for control messages to detect joins and exits, and uses heartbeat and probe messages to identify failures:
\begin{itemize}
    \item \textit{Node joins:} When a new node joins, it sends a \textit{scale-out} message to the scheduler with its IP-port and available links, then Algorithm \ref{alg:FeasibleAllocation} is invoked.
    \item \textit{Node exits:} When a node leaves, it sends a \textit{scale-in} message to the scheduler.
    \item \textit{Node failure:} Chaos uses heartbeats to track liveness. Each node sends a heartbeat to the scheduler every few seconds. If no heartbeat arrives within a set time, the node is considered lost, and \textit{scale-in} is triggered.
    \item \textit{Link joins:} To create a new link, a node sends a \textit{connect-link} message to the scheduler, specifying the target node.
    \item \textit{Link exits:} To remove a link, a node sends a \textit{disconnect-link} message to the scheduler, specifying the target node.
    \item \textit{Link failure:} Link failures are detected using probes. If two nodes fail to measure bandwidth or latency on a link, it is considered down, and one end sends a \textit{disconnect-link} message to the scheduler.
\end{itemize}

Table \ref{tab:commands} summarizes the scaling primitives used in Chaos. Scale-out causes the most delay due to the need to synchronize training state, while others involve only light communication delays that are negligible. For their implementation details, please go to Section \ref{sec:peer-negotiation-protocols}.

\textbf{Network Resource Monitoring.}
When a new node joins, Chaos invokes Algorithm \ref{alg:FeasibleAllocation}, which relies on propagation delay $t_{u\rightarrow v_{\text{new}}}^{\text{prop}}$, transmission delay $t_{u\rightarrow v_{\text{new}}}^{\text{trans}}$, and synchronization delay $\tau_u^{\text{sync}}$. At this point, network resource monitoring is activated. Propagation and transmission delays are measured using iperf, while synchronization delay is measured with timestamps, one recorded before and one after all-reduce. Their difference gives the synchronization delay. Since resource monitoring is expensive, Chaos measures these metrics asynchronously and only when necessary.

\subsection{Peer Negotiation Protocols}\label{sec:peer-negotiation-protocols}
Fig. \ref{fig:peer-negotiation-protocols} shows how nodes coordinate for the scaling primitives in Table \ref{tab:commands}. In case 1, a node joins and then leaves; in case 2, one link is added and another is removed.

\textbf{Scale-out.} As shown in Figs. \ref{fig:peer-negotiation-protocols}a and \ref{fig:peer-negotiation-protocols}e, once the new node 4 finishes initialization (e.g., preparing training data, building the model, allocating RAM and VRAM), it sends a join request to the scheduler, including its IP and port, and the IDs of intended neighbors (e.g., nodes 1, 2 in case 1). According to the negotiation plan, the scheduler first instructs the neighbors to establish socket connections with the new node. It then triggers the cluster monitor to gather bandwidth and latency data, and runs Algorithm \ref{alg:FeasibleAllocation} to generate a state replication plan. Since the topology has changed, the model synchronization policy should also be updated. After all-reduce is completed for the current iteration, the scheduler sends the state replication policy to the new node and its neighbors, and the model synchronization policy to all nodes. According to the replication policy, neighbor 1 sends its training state shards in set $\mathcal{K}_1$ to the new node, while neighbor 2 sends the shards in $\mathcal{K}_2$. These steps run asynchronously with gradient computation, so their delays can overlap. Once state replication and gradient computation finish, the node proceeds with all-reduce using the new model synchronization policy.

\textbf{Scale-in.} As shown in Figs. \ref{fig:peer-negotiation-protocols}b and \ref{fig:peer-negotiation-protocols}e, after node 4 joined, we now remove it. To exit, node 4 sends a leave request to the scheduler. With the scheduler's coordination, it disconnects from neighbor nodes 1 and 2. 
Since the network topology has changed, the scheduler must also update the model synchronization policy accordingly. All nodes then proceed with all-reduce using the new policy. In a node exit, operations such as disconnecting links and updating model synchronization policy happen after all-reduce. But for a node failure, they should be performed before all-reduce to avoid networking errors. If the failure happens during all-reduce, the scheduler notifies all nodes to restart all-reduce. Besides, a node failure is triggered when the scheduler misses heartbeats from node 4 for too long, then the scale-in negotiation will start automatically. The rest follows the same steps as a normal exit. The scheduler will periodically back up its state with neighbors, so that if the scheduler crashes, another scheduler can take over.

\textbf{Connect-link.} As shown in Figs. \ref{fig:peer-negotiation-protocols}c and \ref{fig:peer-negotiation-protocols}f, we add a new link between nodes 1 and 4. In this case, node 4 sends a link join request to the scheduler, asking to connect to node 1. The scheduler starts peer negotiation and instructs both nodes to establish the socket. Once connected, the cluster monitor gathers network data and updates the model synchronization policy. Since the cluster monitor needs an active socket to measure bandwidth and latency, monitoring should happen after the link is up. While network measurement adds some latency, it overlaps with all-reduce and gradient computation, so its delay can be hidden. After the current all-reduce completes, the scheduler notifies all nodes to update their model synchronization policy and continue with this new setting.

\begin{figure}[t]
\centerline{\includegraphics[width=.9\linewidth]{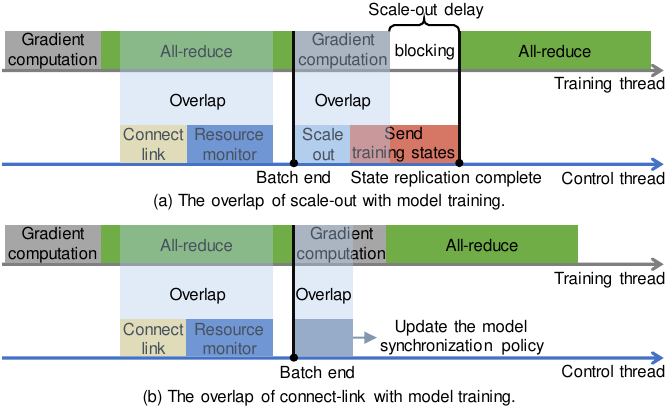}}
\caption{The overlap between (a) scale-out, (b) connect-link with model training.}
\label{fig:overlap}
\end{figure}

\textbf{Disconnect-link.} As shown in Figs. \ref{fig:peer-negotiation-protocols}d and \ref{fig:peer-negotiation-protocols}f, now we remove the link between nodes 2 and 4. In this case, node 4 sends a link leave request to the scheduler, asking to disconnect from node 2. Once the current all-reduce finishes, nodes 2 and 4 disconnect from each other under the scheduler’s coordination. Then, each node updates its model synchronization policy to adapt to the new topology. The disconnect-link delay is negligible because it overlaps with gradient computation. If a link fails, e.g., all-reduce or iperf crashes between nodes 2 and 4, node 4 sends a link leave request to the scheduler. The scheduler then instructs all nodes to update the model synchronization policy and restart all-reduce.

\subsection{Overlap Analysis}\label{sec:overlap}
Chaos achieves the low latency not only due to its efficient state replication, but also thanks to the overlap between model training and peer negotiation. In Fig. \ref{fig:overlap}a, when a new node joins during all-reduce, the control thread handles socket setup and resource monitoring in parallel. Although resource monitoring incurs some delay, this delay is largely hidden by all-reduce. After that, state replication overlaps with gradient computation, allowing a part of the replication delay to be hidden. This overlap further reduces scale-out delay beyond what is achieved by multi-neighbor replication with shard assignment. Another example is shown in Fig. \ref{fig:overlap}b, where connect-link has no state replication, so its delay is fully hidden by gradient computation. Scale-in and disconnect-link are even lighter because they don't require resource monitoring. This explains why scale-in, connect-link, and disconnect-link consistently stay under 20ms (see Table \ref{tab:commands} and Fig. \ref{fig: Overhead of Network Link Changes}).
\begin{figure}[t]
\centerline{\includegraphics[width=.85\linewidth]{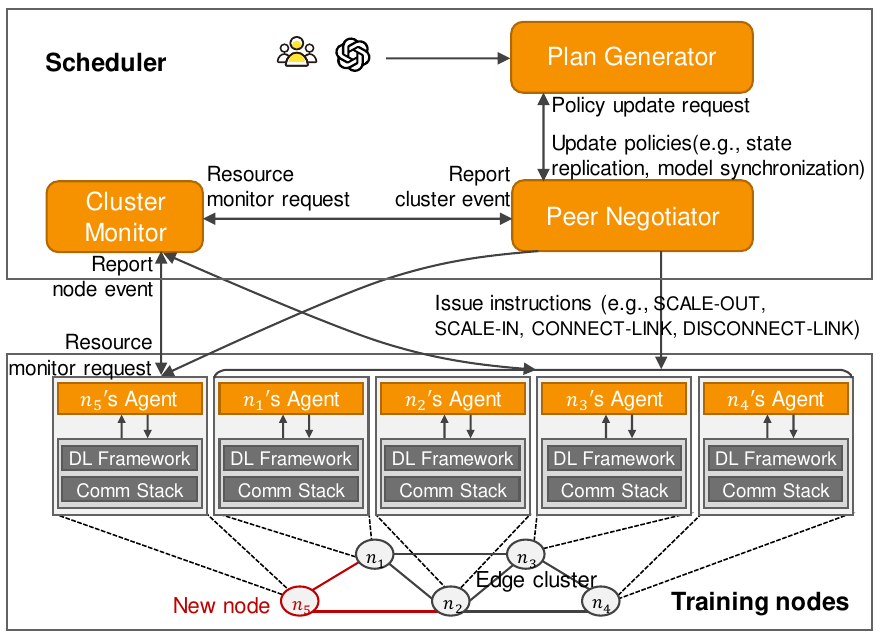}}
\caption{System overview.}
\label{fig:constellation-system-architecture}
\end{figure}

\section{System Design and Integration}\label{system_design}
\subsection{System Design}
Fig. \ref{fig:constellation-system-architecture} illustrates the overall Chaos architecture, consisting of two main node types: training nodes and the scheduler.

\subsubsection{Training nodes} 
These nodes are the core compute units powered by three levels (from down to up): 
\begin{itemize}
    \item \textit{Communication stack} handles data exchange between training nodes, e.g., weights, gradients, activations.
    \item \textit{DL framework} runs the training pipeline, including data preparation, gradient computation, and parameter updates. Frameworks like PyTorch, TensorFlow, and MXNet can be used.
    \item \textit{Agent} bridges the node and scheduler. On the north bridge, it monitors local status, detects node and link changes, and reports events to the scheduler to trigger scaling actions. It also executes commands from the scheduler, such as connecting to a new neighbor, disconnecting a link, sending or receiving training state, and updating local synchronization policy. On the south bridge, the agent coordinates computation and communication. For example, establishing or closing sockets, updating communication rules (e.g., which data to send or receive), synchronizing training state for new nodes, and handling faults like blocking and restarting all-reduce.
\end{itemize}

\subsubsection{Scheduler}
The scheduler is the brain of Chaos and can be deployed on any node. Its \textit{cluster monitor} tracks cluster topology, network resources, and scaling events reported by node agents, then passes this data to the \textit{peer negotiator} to take actions based on the protocols described in Section \ref{sec:peer-negotiation-protocols}. For each event, the scheduler should update the model synchronization policy since the topology has changed, and for scale-out, it also updates the state replication policy. These policies are generated by the \textit{plan generator} using data from the cluster monitor. Algorithm \ref{alg:FeasibleAllocation} works in the control plane and outputs the state replication policy, while the model synchronization policy follows the data-plane all-reduce strategy (e.g., \cite{li2024accelerating}). The plan generator supports custom policies, like user-defined or AI-generated ones.

\begin{figure}[t]
    \centering
    \noindent\rule{\linewidth}{0.8pt}
    \vspace{-1.5mm}
    \begin{lstlisting}[style=custompython,caption=PyTorch code example of node joins and leaves,label=lst:node-join, moredelim={[is][\color{red}\textbf]{@}{@}}]
from chaos import Coordinator
trainer = torch.Trainer()
coordinator = Coordinator() # This will send a join request to the scheduler.
for batch_data in train_loader:
  torch.forward_and_backward()
  trainer.update()  # All-reduce and update
  coordinator.check_state_replication() # Fetch training state from neighbors
  if capture_exit_event():
    coordinator.request_node_exit()  # Send a leave request to the scheduler
\end{lstlisting}
\end{figure}

\begin{figure*}[t]
        \centering
    \includegraphics[width=\linewidth]{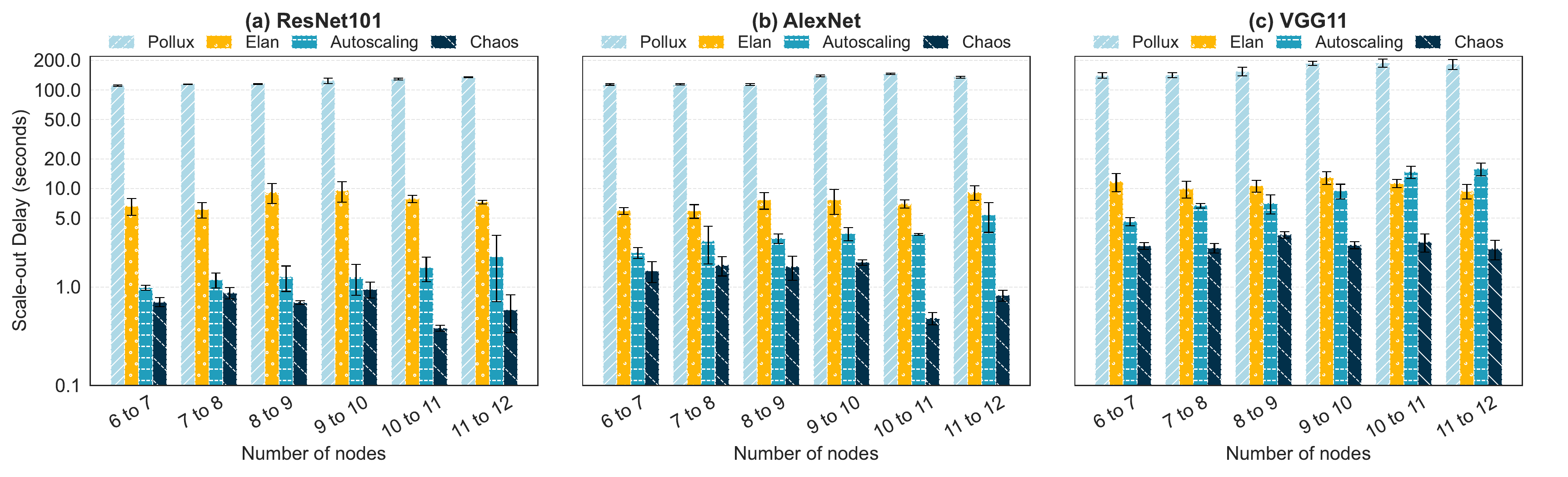}
    \caption{Comparison of scale-out delays of Pollux, Elan, Autoscaling, and Chaos.}
    \label{fig: Scaling Overheads}
\end{figure*}

\subsection{Use Chaos with Common Deep Learning Frameworks}
Chaos is a control-plane system that leaves data preparation, model computation, and all-reduce communication untouched, so it is compatible with any deep learning framework. We designed easy-to-use APIs to integrate Chaos with common DL frameworks:
\begin{itemize}
    \item \texttt{Coordinator}: A class that, when instantiated, sends a node join request to the scheduler. Then, following the scale-out workflow, it establishes socket connections with its neighbors and synchronizes training state. 
    \item \texttt{check\_state\_replication}: This function runs on existing nodes to check if they need to synchronize training state. If the scheduler issues a state replication policy, neighbors should send their model shards to the new node following this policy.
    \item \texttt{capture\_exit\_event:} A helper function that detects if a user node intends to leave. For example, when the user clicks ``Quit Cluster" in the GUI or presses ``Ctrl+C" in the terminal.
    \item \texttt{request\_node\_exit}: Send a leave request to the scheduler to initiate the scale-in workflow and quit the cluster safely. It will disconnect from its neighbors.
\end{itemize}

Listing \ref{lst:node-join} shows an example code for node joins and leaves using PyTorch. For node joining, on the new node, the user initiates the scale-out workflow by creating a \texttt{Coordinator()} instance, and others call \texttt{check\_state\_replication()} after all-reduce to check if they need to synchronize training state to the new node. For node leaving, the node calls \texttt{capture\_exit\_event()} to check if its owner intends to quit. If so, it initiates the scale-in workflow by calling \texttt{request\_node\_exit()}. In these cases, connect-link and disconnect-link are also triggered, since adding a node involves creating new sockets, while leaving removes them.

For node and link failures, they require no extra code. Chaos detects them automatically via heartbeats and probes. Once detected, the scheduler autonomously triggers the peer negotiation protocol to handle them.
\section{Performance Evaluation}
\label{sec_implementation}
\subsection{Experimental Setup}
\textbf{Testbed setup.} We developed the Chaos prototype on PS-lite with 2K lines of code. The hardware platform comprises a server with 2 Intel Xeon Gold 5220R CPUs (96 cores in total), 754 GB RAM, and 8 NVIDIA GeForce RTX 3090 GPUs. We used Docker to virtualize 6-12 nodes, each with 1 GPU card. Since Chaos operates in the control plane, its latency measurement is independent of GPU type and count. To simulate node changes, new nodes join by randomly connecting to several neighbors. Node exits or failures are triggered by randomly removing a node. For link changes, we randomly connect two nodes to add a link, or disconnect a random link to simulate an exit or a failure. To emulate WAN dynamics and heterogeneity, we used Linux tc to randomly set bandwidth (100-1000Mbps) and link latency (10-20ms), updating the settings every 3 minutes, following the configuration in \cite{li2024accelerating}.

\textbf{Training task.} We set up three types of training tasks: (a) Image classification on CIFAR-10 using ResNet101, AlexNet, and VGG11 (ranging from 178 MiB to 528 MiB); (b) Text pre-training on WikiText-2 using GPT-2, GPT-2 medium, and GPT-2 large (ranging from 468 MiB to 3050 MiB); (c) GPT-2 small fine-tuning with LoRA (1.7 MiB) on DailyDialog.

\textbf{Performance metrics.} We focus on three metrics: how node joins/leaves affect convergence accuracy; the delay caused by adding/removing nodes and links; and the time wasted waiting.
\begin{itemize}
    \item \textit{Accuracy:} When nodes join or leave, they bring in or take away their training data. A change in training data can affect convergence.
    \item \textit{Scaling delay:} Chaos implements four scaling primitives: scale-in, scale-out, connect-link, and disconnect-link. We measure the time each primitive will take.
    \item \textit{Cluster idle time:} During scale-out, some nodes may pause for state replication if not fully overlapped with model computation, leaving their GPUs idle. We measure the total GPU idle time across the cluster.
\end{itemize}

\begin{figure*}[t]
    \centering
    \includegraphics[width=\linewidth]{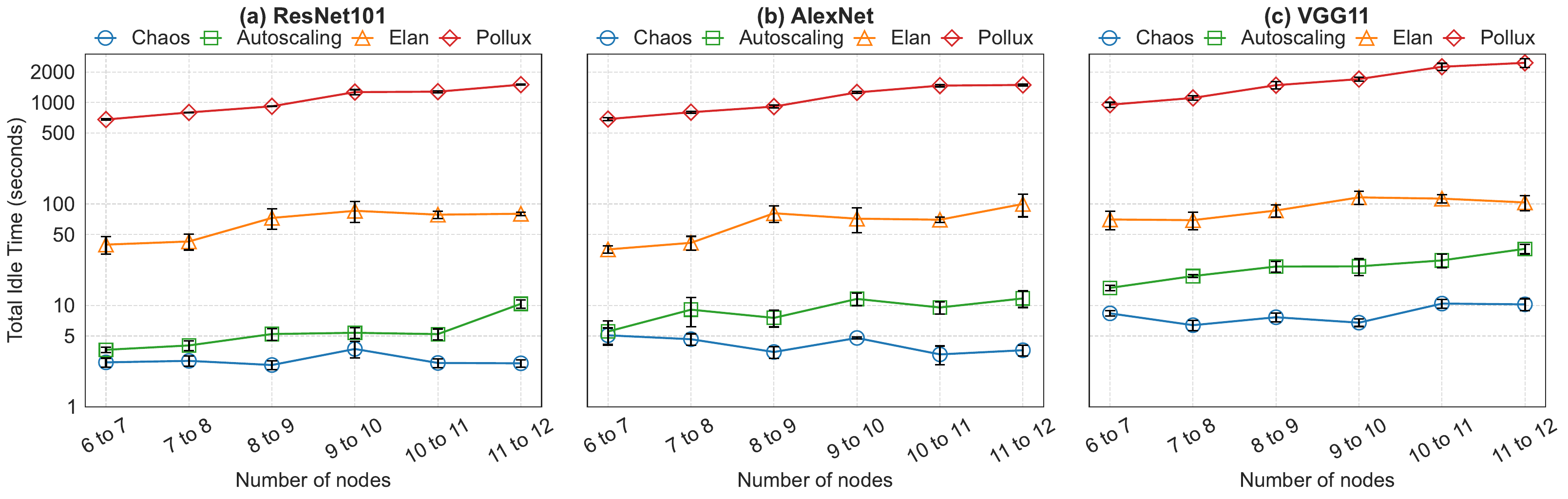}
    \caption{The total GPU idle time across the cluster during each scale-out.}
    \label{fig: Cluster idle time.}
\end{figure*}

\textbf{Benchmarks.} Among the four scaling primitives, \textit{scale-out} is the most time-consuming due to state replication, so \textit{we focus on its delay and compare it with the scale-out delay in Pollux \cite{qiao2021pollux}, Elan \cite{xie2020elan}, and Autoscaling \cite{or2020resource}.}
\begin{itemize}
    \item \textit{Pollux}: Use checkpoint-restart in scaling actions, which will stop training and restart afterwards. 
    \item \textit{Elan}: Use stop-free autoscaling with single-source replication, which selects the fastest neighbor for state replication. Fig. \ref{fig:Motivation1}b shows an example.
    \item \textit{Autoscaling}: Use stop-free autoscaling with multi-source replication, allowing the new node to fetch from multiple nodes in parallel. Fig. \ref{fig:Motivation1}c shows an example.
    \item \textit{Chaos (ours):} Use stop-free autoscaling with multi-neighbor replication, where the new node fetches from multiple neighbors with smart shard scheduling. Fig. \ref{fig:Motivation1}d shows an example.
\end{itemize}

% \textit{(b) Ablation Study:} To evaluate the efficiency of multi-neighbor replication proposed in Section \ref{sec:motivation}, we use single-source replication in Elan and multi-source replication in Autoscaling as benchmarks.
% \begin{itemize}
%     \item \textit{Single-source replication:} The new node pulls the full training state from one fastest neighbor.
%     \item \textit{Multi-source replication:} The new node pulls different shards of the training state from several nodes, with even shard scheduling.
%     \item \textit{Multi-neighbor replication (ours):} The new node pulls different shards of the training state from several neighbors, with even shard scheduling.
% \end{itemize}

\subsection{Comparison of Scale-out Delays for Pollux, Elan, Autoscaling, and Chaos}
This experiment compares the scale-out delay of Pollux, Elan, Autoscaling, and our Chaos. The results are shown in Fig. \ref{fig: Scaling Overheads}, each run was repeated 5 times. The x-axis ``6 to 7" means one node was added to a 6-node cluster during training, and others follow the same pattern. Since Chaos's scale-out delay is so small, we use a logarithmic scale to make it visible.

Different scaling strategies lead to significant differences in scale-out delays. Pollux has the highest delay, over 100 seconds, mainly due to checkpoint I/O, which requires writing and reading the training state to and from a slow disk. Elan, Autoscaling, and Chaos avoid disk access and instead use the network to pull the training state, thus having much lower delays. However, Elan still faces delays above 10s for larger models because it transfers the entire training state from a single node to the new node, creating a bottleneck as model size grows. Autoscaling is faster than Elan in small setups, but its performance drops as the cluster and model scale up, eventually lagging behind Elan. The main issue is redundant traffic: since all nodes share identical training states, multi-hop transfers add no value, they just waste bandwidth. As the model and cluster size grow, this overhead becomes even more pronounced. Chaos, by contrast, fetches the training state from neighbors. This reduces network hops and better utilizes the bandwidth of multiple links on the new node, thus speeding up transmission. In addition, with a smart shard scheduler, Chaos further balances traffic across these parallel links and minimizes the overall replication time. With these optimizations, Chaos cuts the scale-out delay to $\sim$1s. Moreover, we observe that Chaos's scale-out delay grows linearly with model size, as Algorithm \ref{alg:FeasibleAllocation} runs in $O(mK)$, linear in the number of model shards $K$.

\begin{figure}[t]
    \centering
    \includegraphics[width=0.7\linewidth]{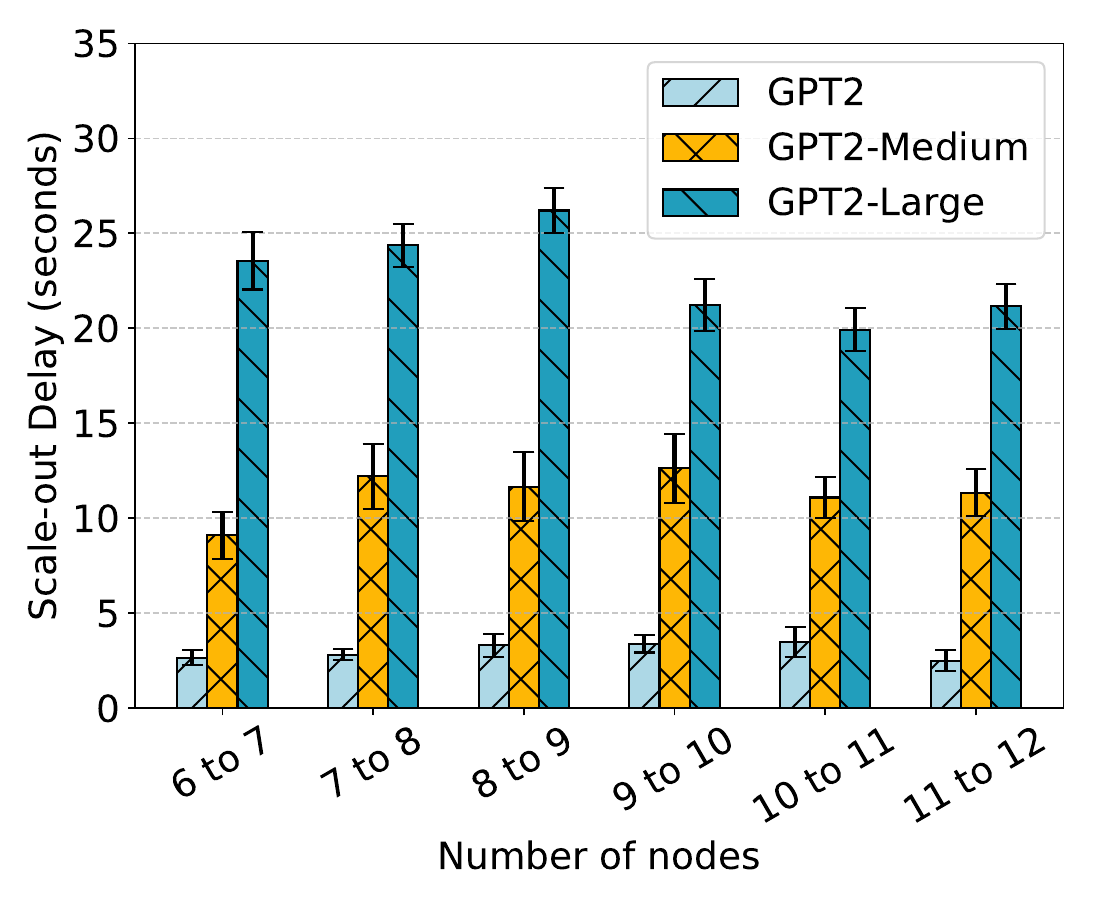}
    \caption{Scale-out delays of Chaos on GPT-2 models.}
    \label{fig: Versatility of the Constellation}
\end{figure} 

\begin{figure}[t]
    \centering
    \includegraphics[width=0.7\linewidth]{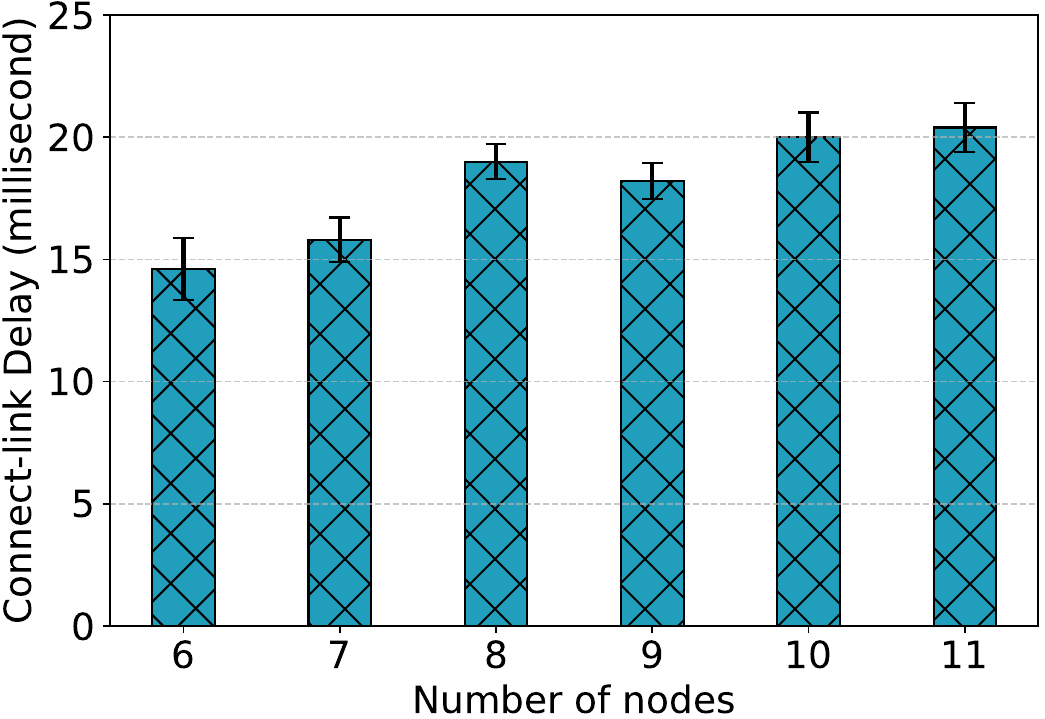}
    \caption{The connect-link delay in Chaos as cluster size grows.}
    \label{fig: Overhead of Network Link Changes}
\end{figure} 

\begin{figure*}[t]
    \centering
    \includegraphics[width=\linewidth]{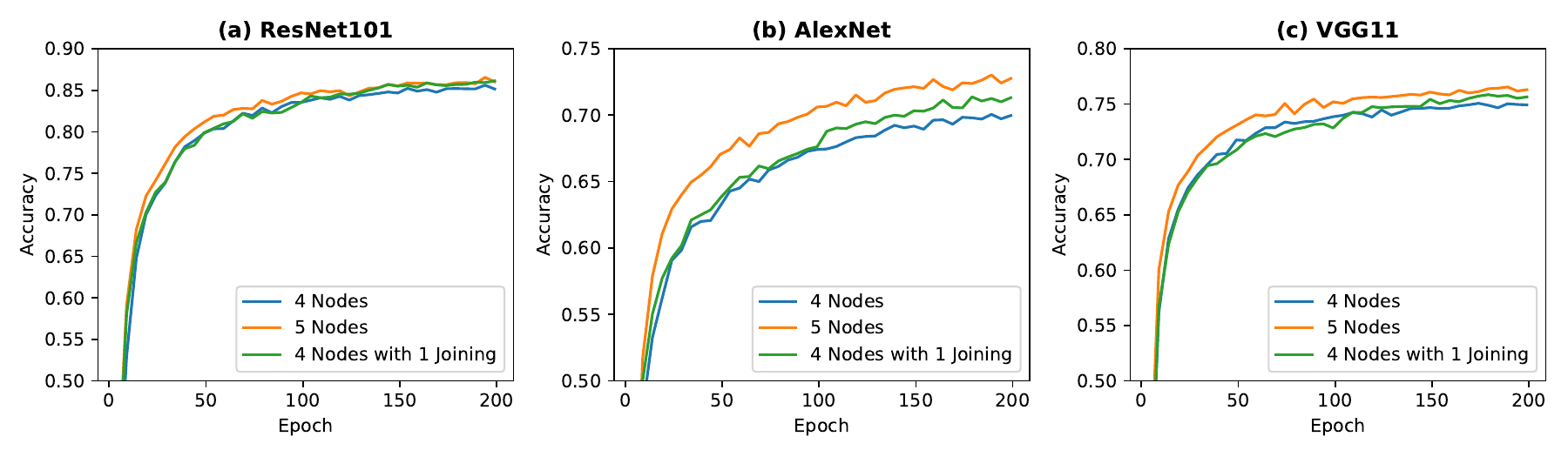}
    \caption{The effects of scale-out on ResNet101, AlexNet, VGG11 convergence.}
    \label{fig: The impact of node addition on model convergence.}
\end{figure*}

\begin{figure*}[t]
    \centering
    \includegraphics[width=\linewidth]{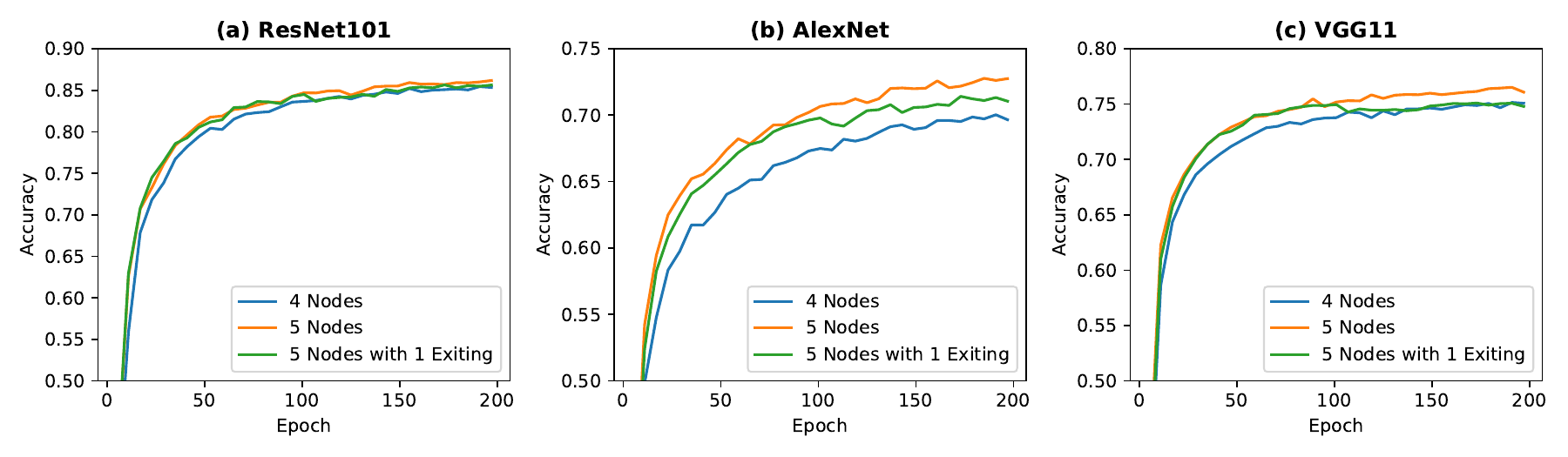}
    \caption{The effects of scale-in on ResNet101, AlexNet, VGG11 convergence.}
    \label{fig: The impact of node exit on model convergence}
\end{figure*}

We also find that Chaos offers superior scalability. As the cluster grows, delays increase for Pollux, Elan, and Autoscaling, but Chaos’s delay stays flat, and even drops when the cluster grows beyond 10 nodes. This is because a larger cluster gives the new node more neighbors and greater parallel bandwidth, thus speeding up state transfer. This advantage is unique to Chaos; Elan and Autoscaling cannot achieve it, since bigger networks mean longer transfer paths and more chances for bandwidth bottlenecks. Results on text generation tasks with GPT-2 models confirm this as well. As shown in Fig. \ref{fig: Versatility of the Constellation}, as model size grows, scale-out delays increase linearly, but stay stable as the cluster gets larger.

\subsection{Comparison of Cluster Idle Time of Pollux, Elan, Autoscaling, and Chaos}
Cluster idle time is the total GPU idle time of nodes during scale-out. It reflects how much computing resources are wasted. A system might have a high scale-out delay but affect only one node, making total idle time low; or have a low scale-out delay but affect all nodes, making total idle time high. Fig. \ref{fig: Cluster idle time.} compares the cluster idle time across Chaos, Autoscaling, Elan, and Pollux. Among them, Pollux wastes the most resources, with cluster idle time exceeding 1000 seconds, because its high checkpoint-restart delay affects all nodes. Next is Elan, with idle time between 50 and 100 seconds. Although the new node fetches from only one node, an extra barrier in Elan forces all nodes to wait, making the total idle time high. Autoscaling also shows high idle time because many nodes participate in state replication, and its scale-out delay grows quickly as the cluster expands. In contrast, Chaos consistently keeps cluster idle time under 10 seconds thanks to its fast scale-out and efficient design: only a few neighbors handle state replication while the rest continue training. With scalable scale-out delays and a limited number of neighbors, Chaos keeps idle time low and unaffected by cluster size, showing excellent scalability and resource efficiency.

\subsection{Scale-in, Connect- and Disconnect-link Delays in Chaos}
In the above, we focus on scale-out because it requires state replication and has noticeable communication delays. However, other scaling primitives, including scale-in, connect-link, and disconnect-link, exchange only tiny control messages, and their delay overlaps with heavy operations like all-reduce and gradient computation, making them negligible. In this experiment, we add a link between two randomly chosen nodes. As shown in Fig. \ref{fig: Overhead of Network Link Changes}, the connect-link delay stays below 20ms, so it's unnecessary to compare with baselines since it cannot become a bottleneck. This result also applies to scale-in and disconnect-link, because they have a similar traffic model.

\begin{figure}[t]
    \centering
    \begin{subfigure}[b]{0.49\linewidth}
        \centering
        \includegraphics[width=0.99\linewidth]{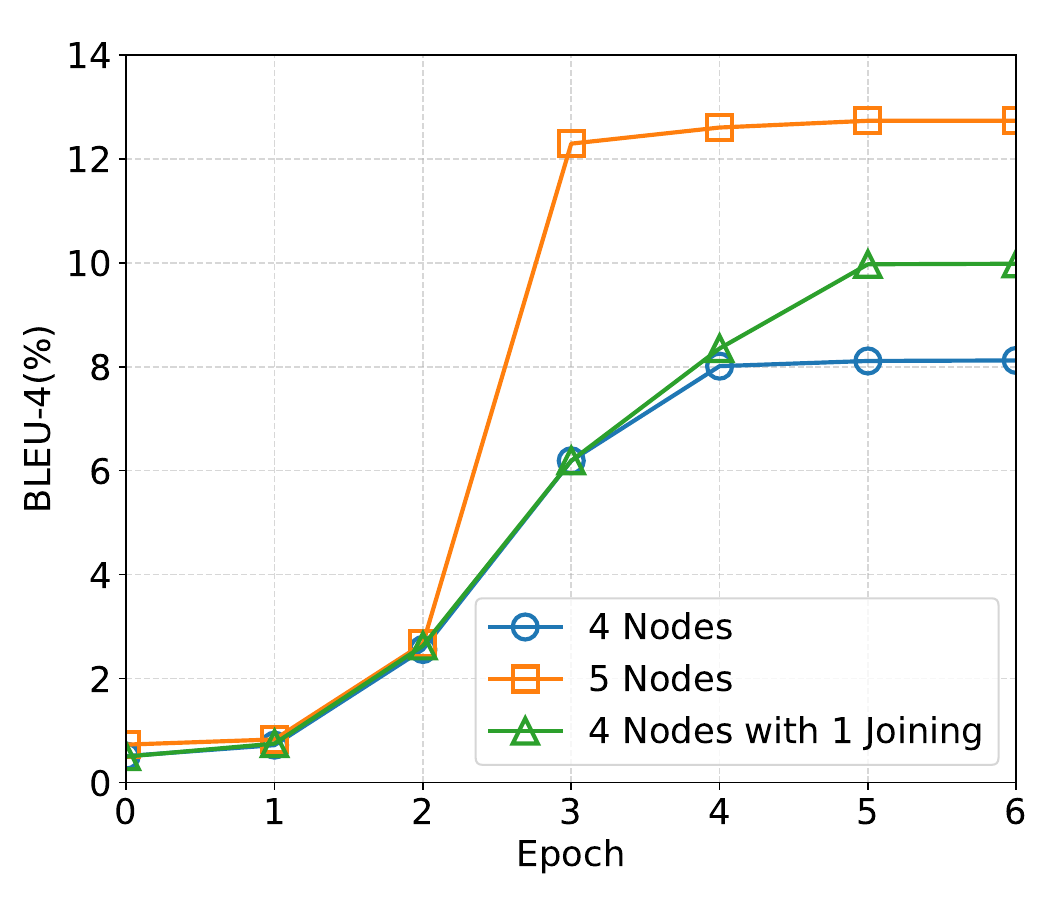}
        \caption{Scale-out}
    \end{subfigure}
    \hfill
    \begin{subfigure}[b]{0.49\linewidth}
        \centering
        \includegraphics[width=0.99\linewidth]{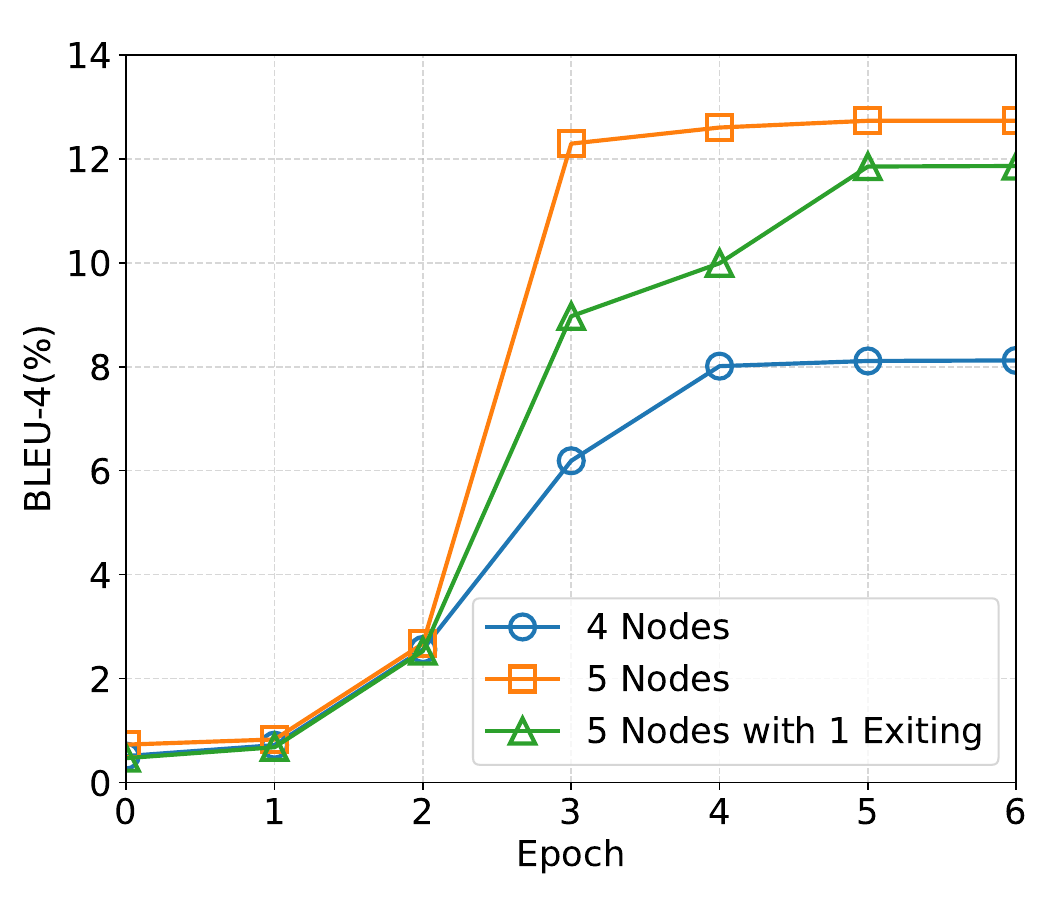}
        \caption{Scale-in}
    \end{subfigure}
    \caption{Effects of scale-out, scale-in on GPT-2 convergence.}
    \label{fig:gpt2-scale-converge}
\end{figure}

% \begin{figure}[t]
%     \centering
%     \includegraphics[width=.7\linewidth]{figures/exp/The-impact-of-node-addition-on-BLUE4-GPT2.pdf}
%     \caption{The effects of scale-out on GPT-2 convergence.}
%     \label{fig: The effects of scale-out on GPT-2 BLEU-4 Convergence}
% \end{figure} 

% \begin{figure}[h]
%     \centering
%     \includegraphics[width=.7\linewidth]{figures/exp/The-impact-of-node-exiting-on-BLUE4-GPT2.pdf}
%     \caption{The effects of scale-in on GPT-2 convergence.}
%     \label{fig: The-impact-of-node-exiting-on-BLUE4-GPT2}
% \end{figure} 

\subsection{Ablation on How Scale-in and Out Affect Convergence?}
Adding or removing nodes changes the training data, so this experiment assesses how scale-out and scale-in affect model convergence. The accuracy curves for ResNet101, AlexNet, and VGG11 are shown in Figs. \ref{fig: The impact of node addition on model convergence.} and \ref{fig: The impact of node exit on model convergence}, respectively. For scale-out, training starts with 4 nodes, and a new node is added at the 100th epoch (green). Before the new node joins, the scale-out curve (green) closely tracks the 4-node curve (blue), and it smoothly shifts toward the 5-node curve (orange) after the new node joins. For scale-in, training starts with 5 nodes, and one node is removed at the 100th epoch (green). Before a node leaves, the scale-in curve follows the 5-node curve and gradually shifts toward the 4-node curve afterward. Even after the node leaves, the model retains knowledge from its data. The same pattern was found when fine-tuning GPT-2 with LoRA, as shown in Fig. \ref{fig:gpt2-scale-converge}. This shows that Chaos also works well for LLM fine-tuning. Moreover, both scale-out and scale-in curves remain smooth during node churn, without sharp fluctuations, because deep models are robust to small changes in training data, and minor additions or removals do not disrupt convergence. Since we change only one node at a time, the noise is further diluted by global averaging. 

\begin{figure}[t]
    \centering
    \includegraphics[width=.7\linewidth]{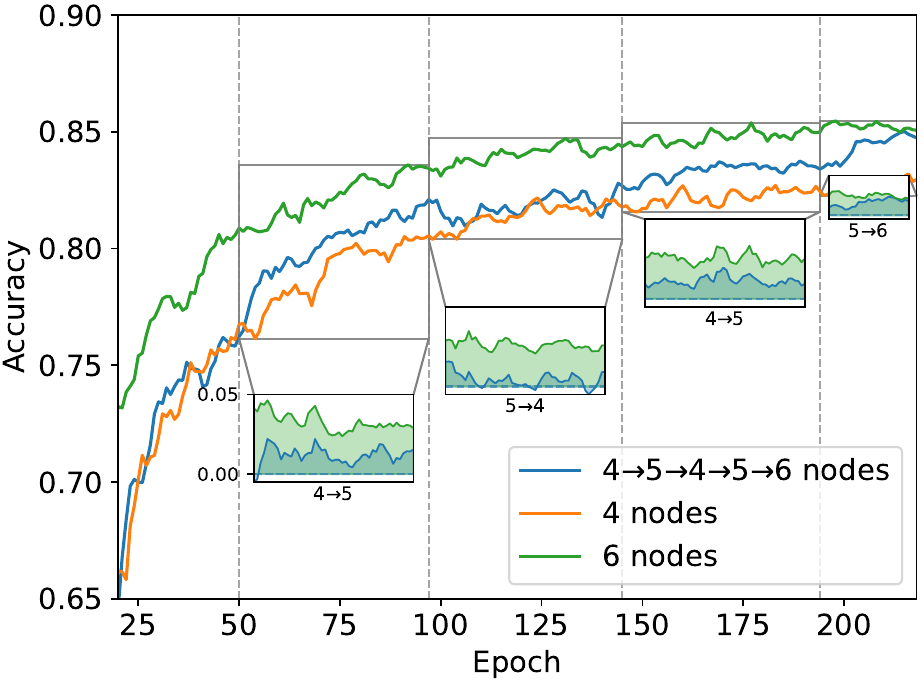}
    \caption{Global accuracy curves on ResNet101. Green: fixed 6 nodes. Orange: fixed 4 nodes. Blue: 4-6 nodes with churn.}
    \label{fig:global-acc}
\end{figure}

Fig. \ref{fig:global-acc} shows the global accuracy curves on ResNet101. Training starts with 4 nodes and undergoes four churn events: one node joins at epoch 50; one node exits at epoch 97; one node joins at epoch 145; and one more node joins at epoch 194, ending with 6 nodes. The sub-figures plot accuracy gains over the 4-node baseline. After each join, accuracy steps up by a few points, reflecting the benefit of more parties (data). The node's leave causes a small dip, but the model retains knowledge learned from the exited node, as the overall gain remains positive. These align with the local trend in earlier experiments. From a global view, training under churn (blue) converges to nearly the same accuracy as the 6-node run (green). This suggests that multi-party training under churn is feasible. We focus on exploring the effect of node churn on model accuracy, but methods for preserving memory under churn are complementary and left for future work. Nevertheless, prior studies like \cite{sun2023mimic} have already shown promising solutions to mitigate forgetting in such settings.

\subsection{Ablation on State Replication Mechanism}
State replication is key to a fast scale-out. In this experiment, we run an ablation study to evaluate our replication mechanism, highlighting the benefits of multi-neighbor replication and our greedy sharding algorithm. The single-source method fetches from the fastest neighbor, whereas the multi-source method fetches from multiple nodes (not neighbors) with the model evenly sharded across them. In addition to our multi-neighbor replication with greedy sharding, we also provide a variant with equal sharding to observe the individual benefits of multi-neighbor replication and greedy sharding.

\begin{figure}[t]
    \centering
    \includegraphics[width=.7\linewidth]{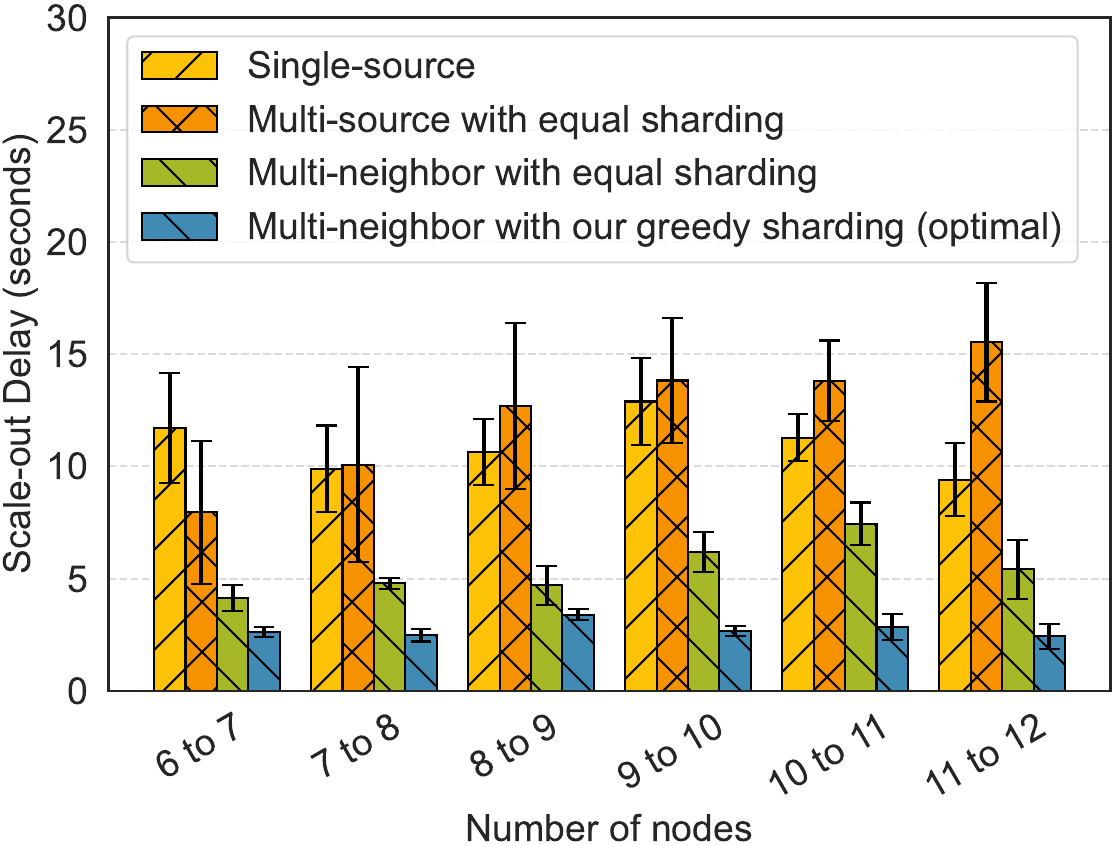}
    \caption{Comparison of single-source, multi-source, and multi-neighbor state replication.}
    \label{fig: Ablation}
\end{figure}

As shown in Fig. \ref{fig: Ablation}, multi-neighbor methods achieve lower scale-out delays than single- and multi-source methods, with multi-source often even slower than single-source. This aligns with Fig. \ref{fig:Motivation1}: single-source suffers from bandwidth and load bottlenecks that worsen as models and clusters grow, while multi-source adds redundant traffic and multi-hop forwarding, leading to rapidly increasing delays. In contrast, multi-neighbor replication fetches only from neighbors, eliminating redundancy and long paths. With greedy sharding, Chaos keeps scale-out delays lower and more stable, and even faster as the cluster grows, by precisely balancing transmission loads across neighbors. Its sharding and assignment solution has been proven optimal in Section \ref{sec_state_replication_mechanism}.

\section{Conclusion}
This paper proposes Chaos, a multi-party distributed training system with self-healing and autoscaling, enabling robust and elastic training under churn. It implements four scaling primitives: scale-out, scale-in, connect-link, and disconnect-link, to handle node and link joins, exits, and failures. To speed up autoscaling, we propose multi-neighbor replication with an optimal sharding and assignment mechanism to minimize scale-out delay. A built-in cluster monitor tracks resource and topology changes, feeding real-time network data to the scheduler for scaling decisions. Through peer negotiation protocols, Chaos handles scaling events in a fully self-governed manner. Experiments show that Chaos achieves substantially lower scale-out delay than Pollux, Elan, and Autoscaling, and handles scale-in, connect-link, and disconnect-link events within 20ms, making it smooth to handle the churn in a cross-region cluster. It also delivers the lowest idle time, showing superior resource use and scalability as the cluster grows.

\section*{Acknowledgments}
We thank SAIBIT TECH for providing equipment support.

\bibliographystyle{IEEEtran}
\bibliography{references}
\balance
\end{document}